\documentclass[sigconf]{acmart}

\usepackage{graphicx}
\usepackage{amsmath}
\usepackage{algorithm}
\usepackage{algpseudocode}

\newtheorem{definition}{\sc Definition}[section]

\AtBeginDocument{%
  \providecommand\BibTeX{{%
    \normalfont B\kern-0.5em{\scshape i\kern-0.25em b}\kern-0.8em\TeX}}}

\begin{document}

\title{Private Approximate Query over Horizontal Data Federation}\titlenote{This work is funded by DigiTrust (\url{http://lue.univ-lorraine.fr/fr/article/digitrust/}).}

\author{Ala Eddine Laouir and Abdessamad Imine}

\affiliation{
  \institution{Universit\'e de Lorraine, CNRS and Inria, Nancy}
  \country{France}
  }
\email{firstname.lastname@loria.fr}

%
%
%
%
%

\begin{abstract}
In many real-world scenarios, multiple data providers need to collaboratively perform analysis of their private data. The challenges of these applications, especially at the big data scale, are time and resource efficiency as well as end-to-end privacy with minimal loss of accuracy. Existing approaches rely primarily on cryptography, which improves privacy, but at the expense of query response time. However, current big data analytics frameworks require fast and accurate responses to large-scale queries, making cryptography-based solutions less suitable.
In this work, we address the problem of combining \emph{Approximate Query Processing} (AQP) and \emph{Differential Privacy} (DP) in a private federated environment answering range queries on horizontally partitioned multidimensional data. We propose a new approach that considers a data distribution-aware online sampling technique to accelerate the execution of range queries and ensure end-to-end data privacy during and after analysis with minimal loss in accuracy.
Through empirical evaluation, we show that our solution is able of providing up to $8$ times faster processing than the basic non-secure solution while maintaining accuracy, formal privacy guarantees and resilience to learning-based attacks.
\end{abstract}

\keywords{Differential privacy, Big Data, Federated data, Online sampling.}


\maketitle

\sloppy

\section{Introduction}
The extensive reliance of individuals on software solutions in daily and professional life has led to an exponential growth of data collected by companies, corporations, government organisations, and even hospitals. These vast mines of data, if carefully and efficiently analysed, can provide valuable insights that guide decision-making and business development. In large-scale studies and research, the analysis must be conducted on several data sources to obtain meaningful conclusions. An example of such a case is during a pandemic, where many hospitals jointly conduct studies to have a global view of the problem.

One of the most commonly used tools to analyse and explore these huge volumes of data is OLAP tasks, where various aggregation queries 
(\verb+SUM+, \verb+COUNT+, etc.) can be issued to learn existing patterns and trends within the data. These aggregation queries may seem simple, but they are very time-consuming in big databases. The analysis of data from multiple data providers comes with two main challenges: privacy and resource/time efficiency.
The privacy issue arises from the fact that this data is personal and sensitive to individuals, and sharing it with other parties can be very harmful. Many regulations and restrictions like GDPR are imposed by governments on how to process and share such sensitive data. 
In the case of a federated environment, where a joint study requires the collaboration of many data providers, data sharing is highly  restricted. Each data provider must ensure the security and privacy of the data collected from their users during and after the analysis.

To satisfy the requirement of end-to-end privacy, many solutions have been proposed in the literature, and most of them rely on cryptography to ensure there is no data leakage during the exchange and query evaluation. 
Secure multiparty computation (SMC) solutions\cite{bater2018shrinkwrap,bater2020saqe} appear to be a prominent 
solution in federated environments. Others use oblivious operations\cite{bater2017smcql} or secure hardware \cite{opaque,oblidb,doquet} so that during query evaluation, each data provider can maintain the confidentiality of their data. Additionally for securing the end result of any OLAP query, Differential Privacy (DP) \cite{dp} is generally considered the gold standard by government and private institutions \cite{apple,census,google1,google2}. 
Due to its strong formal confidentiality guarantees, DP allows individuals to deny their participation in the database. 
These query evaluation solutions in a federated environment meet end-to-end security and privacy requirements.
However, what they have in common is their reliance on encryption. This causes a huge processing time overhead, and for time-sensitive tasks, utility is measured by both accuracy and speed. They certainly address the privacy issue, but they are time and resource consuming.

The issue of reducing query response time has been widely addressed in the literature, through the need to obtain Approximate Query
Processing (AQP). Existing AQP methods can be classified into two types, online approximation and offline synopsis creation. In online approximation, there is \textit{Online Aggregation} based solutions \cite{hellerstein1997online,li2016wander,qin2014pf} that provide fast and reliable approximation of the query continuously, and other solutions based on applying \textit{online sampling} to reduce the processed data and obtain an approximation from a sample \cite{sapprox,goiri2015approxhadoop,song2018approximate}.In offline synopsis creation, views are generated offline using query workloads or/and data statistics \cite{acharya1999aqua,agarwal2013blinkdb,chaudhuri2007optimized}.\\
In this area of research, the main focus is on efficiency, but privacy has not been considered.


In our work, we address the challenge of answering OLAP aggregation range queries in a federated environment, while preserving end-to-end privacy and improving resource and time consumption for query processing.
Our solution relies heavily on differential privacy to secure collaboration and end results, and ensure no information leaks.
To speed up queries, we implement a cluster-based sampling method using a well-known statistical estimator that provides accurate estimates for range queries (such as \verb+SUM+ and \verb+COUNT+) while processing minimal data portions. 
While existing systems ensure either privacy or speedup for query approximation,  to the best of our knowledge, our solution is the first to offer speedup over plain-text execution with end-to-end privacy in a federated environment. Our main contributions can be listed as follows:

\begin{enumerate}
    \item Definition of a lightweight collaboration method that determines optimal sampling decisions for data providers to maximize accuracy without needing access to their full datasets or information leakage.
    \item Introduction of data distribution-aware cluster sampling method with DP guarantees for individual privacy.
    \item Meticulous integration of DP at every step with minimal loss of precision.
    \item Extensive experimentation to empirically validate the performance of our approach in terms of accuracy and time efficiency.
    \item Extensive experimentation to ensure the resilience of our system against learning-based attacks.
\end{enumerate}

\noindent\textbf{Roadmap.} The paper is structured as follows: Section \ref{sec:rw} reviews some existing works. Section \ref{sec:notation} introduces the notions used throughout our paper. Section \ref{sec:problem} gives a detailed description of the problem solved by our approach. Section \ref{sec:solution} presents our proposed solution in detail. The extensive evaluation of our approach is given in Section \ref{sec:eval}. In Section \ref{sec:discussion}, we discuss the limitations/extensions of our solution and we conclude in Section \ref{sec:conclusion} by giving some future works.

\section{Related Works}\label{sec:rw}
Due to the increasing size and distribution of databases, querying and exploring such vast volumes for analytical purposes, quickly and without revealing sensitive information, has become a challenge. Here, we describe the state-of-the-art related to our work.

\noindent\textbf{\textit{Approximate Query Processing (AQP).}} 
As the quality of a query is based on its accuracy and response time, especially for time-sensitive tasks like OLAP \cite{WangJ08} and Business Intelligence (BI), approximating the query offers the best way to strike a balance between these two quality factors.

In the early $1990$s, \cite{hellerstein1997online} proposed a new interactive method for query processing that provides a quick initial answer with a certain error, refining it as processing continues. Other works followed in this direction \cite{xu2008confidence,li2016wander,wu2010continuous,qin2014pf}, each enhancing specific aspects of the method by including support for \textit{group by} or propose parallel and distributed versions. Another research direction focuses on processing a small subset of the original data, thereby reducing query run-time. In \cite{olken1986simple,olken1995random,piatetsky1984accurate,song2018approximate}, uniform row-level random sampling is applied online before query processing. 
Although row-level sampling may improve processing time for complex queries, it can introduce overhead and slow down queries that require a full table scan \cite{bi-ber-dbms} (e.g. Bernoulli sampling). To avoid such overhead, the solutions from \cite{acharya1999aqua,agarwal2013blinkdb,chaudhuri2007optimized} create the samples offline. Cluster sampling,  also referred to as \emph{page-sampling} \cite{bi-ber-dbms}, is utilized to speed-up aggregation queries in big databases. Methods in \cite{goiri2015approxhadoop,sapprox,gapprox} use this sampling in the context of Hadoop Map-Reduce framework \footnote{https://hadoop.apache.org}, as it proves to be fast and I/O efficient compared to row-level sampling.\medskip

\noindent\textbf{\textit{Federated query answering.}}
Data is often distributed across multiple locations (e.g. data providers like hospitals and companies) and the collaboration among all parties is necessary to answer range aggregation queries. But for privacy and security reasons, each data provider cannot disclose their data to third parties.

Some solutions rely on secure hardware modules (i.e. \emph{enclaves}), in which all sensitive code and data are processed. Methods in \cite{agrawal2006sovereign,opaque,oblidb} focus on aggregation queries in this setting, and \cite{opaque,oblidb} use intel's SGX for secure processing. 
These solutions are generally efficient, but their reliance on trusted  hardware and weakness to side-channel attacks constitute a limitation. Recently in \cite{doquet}, the notion of \textit{Differential Obliviousness} was used to mitigate the risk of side channel attacks.`

Other recent works presented Secure Multiparty Computation (SMC) query processing engines \cite{bater2017smcql,bater2018shrinkwrap,bater2020saqe}. 
These engines enable data providers to respond to OLAP queries securely by joining data with end-to-end privacy. 
Differential Privacy (DP) is used to perturb the final results, thereby mitigating any inference attacks based on the results. 
While these solutions incur computational overhead, \cite{bater2020saqe} introduced online random sampling to improve secure computing performance by reducing the size of shared data for query processing. 
In \cite{cao2021atlantic}, sampling is performed offline to create a synopsis to further improve performance. Another solution \cite{secrecy} focused on reducing the cost of SMC operation thus obtaining significant improvement in performances.
All of these SMC (or \emph{enclaves})-based protocols are encryption-based, which prevents them from outperforming plain-text query execution. Even with significant improvements introduces in the past years, on real world big tables they still expensive for real-time queries\cite{secrecy}.

To highlight the scale of this problem, we performed a simulation\footnote{\url{https://github.com/AlaEddineLaouir/Federated-Range-Queries.git}} using a synthetic Adult\cite{adult} horizontally distributed on 4 data providers as a federated environment. We ran a set of random range queries, which are the type of queries we focus on. For the query processing, we considered two solutions using SMC: (i) data providers sharing the rows and collectively evaluating the query; and (ii) evaluating the query locally and only sharing the results and computing the final result.
\begin{figure}[H]
    \centering
    \includegraphics[width=0.75\linewidth]{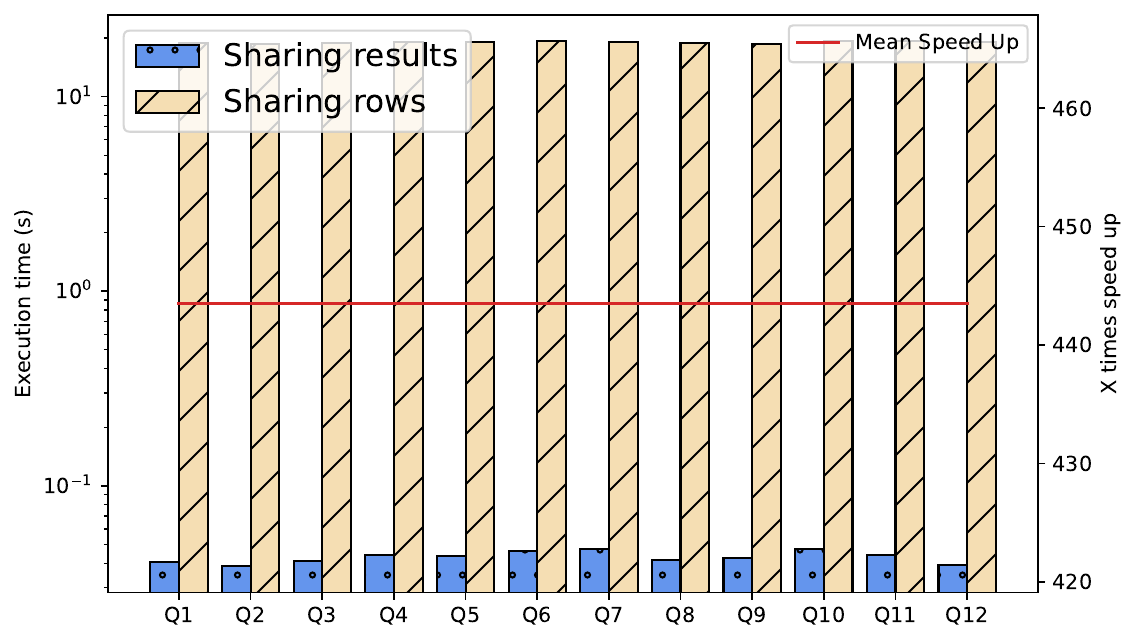}
    \caption{Runtime cost of data sharing in SMC.}
    \label{fig:smc_sim}
\end{figure}

We measured the time required to share the rows/results in SMC. The results in Figure \ref{fig:smc_sim} show that sharing only local results incurs an insignificant overhead of $0.04$ seconds. 
On average, this is less than $440$ times the time required for row sharing in SMC. 
Additionally, the cost of sharing only results remains constant and independent of the dataset, whereas the cost of sharing rows will increase with larger tables.

In our work, we propose a framework to approximate query processing in a federated environment, enabling accelerated query execution compared to plain text execution while ensuring end-to-end Differential Privacy guarantees.


\section{Preliminaries}\label{sec:notation}
In this section, we give the notation and explain briefly notions used throughout the paper. \medskip

\begin{figure}[h]
    \centering
    \includegraphics[width=0.75\linewidth]{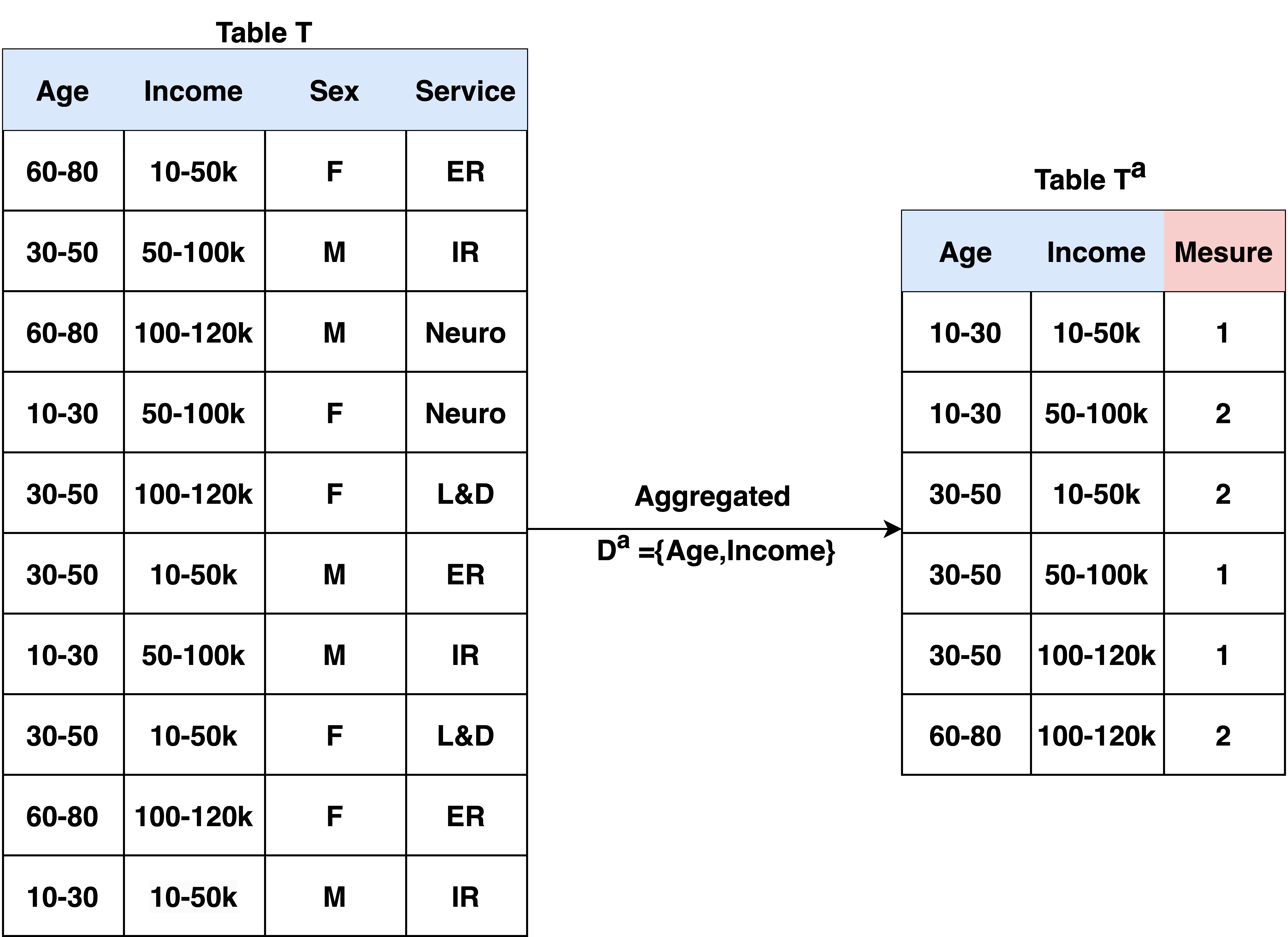}
    
    \caption{Count tensor}
    \label{fig:data_model}
\end{figure}

\noindent\textbf{Data model.} 
In a tabular database $T$ defined over a set of $n$ \emph{dimensions} (or \emph{attributes}) $D=\{d_1,d_2,...,d_n\}$, each individual is a \emph{row} with values on each dimension. We assume that each dimension $d$ is associated with a domain $\vert d \vert$ containing discrete and totally ordered values, the size of the domain is $||d||$. For performance purposes during \textit{online analytics} tasks, the table $T$ is transformed into a \textit{multidimensional data} (or a \textit{count tensor}) $T^a$ of dimensions $ D^{ a } \subset D$, which has an attribute \emph{Measure} storing the number of \emph{aggregated} rows of $T$. Figure \ref{fig:data_model} illustrates how to construct a count tensor $T^a$ from table $T$ by aggregating dimension \textit{Service}.
For simplicity, we use term ``table''  for ``tabular data'' and ``count tensor''.
 
\noindent\textbf{Queries.} To analyze and extract insights from these tables, the analyst can issue aggregation queries, helping to explore the data and gain a general understanding of patterns and trends. In this work, we consider a \emph{range query} $Q$ defined as:

\verb+SELECT Aggregation FROM Table WHERE Range+, where:

\begin{itemize}
    \item \verb+Aggregation+ is \verb+COUNT(*)+ or \verb+SUM(Measure)+.
    \item \verb+Range+ is a set of intervals $r_d = [l_b^d, u_b^d]$ on each dimension $d \in D^Q \text{ where } D^Q \subseteq D$ in \verb+Table+, such that $l_b^d \leq v \leq u_b^d$ for every value $v \in \vert d\vert$.
\end{itemize}

In our work, we focus on COUNT and SUM queries because they are used in several analytics applications. For instance, in a big database aggregating per-stock order data for the NASDAQ exchange, these queries are typically used to analyze order data from past days. Additionally, aggregations, such as average, standard deviation, and variance, can be derived from COUNT and SUM.\medskip

\noindent\textbf{Query Approximation and Sampling.} The goal of query approximation is generally to speed up execution at the expense of answering the query exactly, while preserving answer accuracy as much as possible. 
Online sampling is employed for time-sensitive tasks to reduce the overhead of evaluating queries on large databases. \emph{Note that in this case, the sampling differs from one query to another}.
In statistical terms, random sampling is essentially the process of selecting a subpopulation $SP$ from the total population $P$  where a sampling rate $sr$ dictates the size of $SP$. This subpopulation contains sufficiently representative individuals and properties, capturing various characteristics of $P$ such that the analysis conducted on $SP$ can be generalized to $P$.
All random sampling techniques can be categorized based on three main features:
\begin{itemize}
    \item Granularity: sampling elements are individuals or a bulk/cluster of individuals.
    \item Uniformity: elements are sampled with equal/unequal probabilities.
    \item Replacement: sampling elements can be chosen multiple times or only once.
\end{itemize}

Nowadays, all modern systems choose to split/store a big table $T$ into a set of smaller, manageable entities $T= \{C_1,C_2,... ,C_N\}$ where each entity has a maximum size $S$.
The entity could be \textit{Table pages}\footnote{\url{https://www.postgresql.org/docs/current/storage-page-layout.html}}, \textit{HDFS file Blocks}\footnote{\url{https://hadoop.apache.org/docs/r1.2.1/hdfs_design.html}}, etc.

In this paper, we call these storage entities \textit{Clusters} and we assume that our tables are already stored as a set of clusters.
Given this storage format, sampling on databases can be done at two levels: Row/Cluster level \cite{bi-ber-dbms}.

\emph{In tabular databases with range queries, it is particularly challenging to find an online sampling algorithm that offers speed-up while maintaining accuracy}.

\noindent\textbf{Data providers.} For many real-world use cases, multiple organizations or institutions, called \emph{data providers}, publish access to their databases for joint analysis.
Let $\mathbb{S}$ be the set of data providers. 
In this work, we assume that a large table $T$ is \emph{horizontally} distributed over $\mathbb{S}$ such that all data providers share the same schema (i.e. a set of dimensions) of $T$ but each contains different rows. All data providers use clusters of the same size to store their local tables.
\emph{More importantly, for privacy reasons, data providers collaborate on joint analyzes without revealing their data.}\medskip

\noindent\textbf{Differential Privacy (DP).}  A privacy model that provides formal guarantees of \emph{indistinguishability} such that the query results do not yield much information about the presence or absence of any particular individual. Consequently, it hides information about which of the \textit{neighbouring tables} \cite{dp} was used to answer the query.

\begin{definition}[Neighbouring Tables\cite{dp}]
Two tables $T$ and $T'$ are \emph{neighbouring} if we can obtain one of them by inserting at most a row into the other.
\end{definition}

We use $d(T,T')$ to represent the \emph{distance} between two tables $T$ and $T'$ and we say that two tables are \emph{neighbouring} if their distance is $1$ or less.

\begin{definition}[$(\epsilon,\delta)$-Differential Privacy\cite{dp}]
A mechanism $M$ satisfies $(\epsilon,\delta)$-\emph{Differential Privacy} (or $(\epsilon,\delta)$-\emph{DP}) if, for any two neighboring tables $T$, $T' $ and for any possible output $V$ of $M$:
$$\mbox{Pr}\left[ M\left( T \right) \in  V \right] \leq exp(\epsilon) \times \mbox{Pr}\left[ M\left( T' \right) \in  V \right] + \delta$$
where $\delta$ represents the failure probability. We refer to $(\epsilon,\delta)$ as the \emph{privacy budget}.
\end{definition}

In practice, $M$ is a \emph{randomized} algorithm, which has many possible outputs under the same input.
It is well known that DP is used to answer specific queries on databases. Let $f$ be  a query on a table $T$ whose its answer $f(\mathcal{T})$ returns a number. The \emph{global sensitivity} of $f$ is the amount by which the output of $f$ changes for all neighboring tables.


\begin{definition}[Global Sensitivity\cite{dp}]
For any two neighboring tables $T$ and $T'$, the \emph{global sensitivity} of function $f$ is:
\[ GS_f = \max_{T,T': d(T,T')\leq 1} \left\| f(T) - f(T') \right\|_1 \]
where $\left\| \cdot \right\|_1$ is the $L_1$ norm.
\end{definition}

For instance, if $f$ is a \verb+COUNT+ range query then $GS_f$ is $1$.

The \emph{Laplace Mechanism} is a randomized mechanism for enforcing $\epsilon$-DP (or $(\epsilon,0)$-DP referred to as pure DP), which adds calibrated noise to the output of a function $f$ based on its global sensitivity $GS_f$.

\begin{definition}[Laplace Mechanism \cite{dp}]
The \emph{Laplace Mechanism} adds noise to $f(T)$ as:
\[ S = f(T) + \text{Lap}\left( \frac{GS_f}{\epsilon} \right) \]
where $GS_f$ is the \emph{global sensitivity} of $f$, and $Lap(\alpha)$ denotes sampling from the Laplace distribution with center $0$ and scale $\alpha$.
\end{definition}

Unlike the Laplace Mechanism, which is used to release noisy numerical values, the \emph{Exponential Mechanism} can be used for biased selection of elements from a set based on a scoring function while preserving ($\epsilon,0$)-DP \cite{dp}.

\begin{definition}[Exponential Mechanism \cite{dp}]
Given a set of elements $SE$ and a scoring function $L$, the \emph{Exponential Mechanism} randomly selects $e \in SE$ with the probability of the element $e$ being proportional to:
\[ \exp\left( \frac{\epsilon \times L(e)}{2 \times \Delta_L} \right) \]
where $\Delta_L$ is the sensitivity of $L$.
\end{definition}

\noindent\textbf{Local and Smooth Sensitivity.} In many applications of DP, the global sensitivity $GS_f$ cannot bounded. In this case, there is an alternative definition of sensitivity called \textit{local sensitivity}, where  the maximum difference between the query’s results is based on a fixed database $T$ and any database $T'$ neighbouring to it:
\begin{definition}[Local Sensitivity\cite{nissim2007smooth}]
Given a database $T$ and $T'$ as any of its possible neighbouring tables, the \textit{local sensitivity} of function $f$ is:
\[LS_f(T) = \max_{T':\, d(T, T') \leq 1} \left\| f(T) - f(T') \right\|_1\]
where $\left\| \cdot \right\|_1$ is the $L_1$ norm.
\end{definition}
The local sensitivity $LS_f(T)$ is often much less than the global sensitivity $GS_f$ because it is based on a specific instance of the data $T$. This also makes it unsafe to use, as it can leak information about $T$ on which it is based. Nassim et al \cite{nissim2007smooth}. suggest the use of a smoothing function that finds a safe upper bound for $LS_f(T)$ and can be used to calibrate the randomness (noise) without any risk. These functions usually require that the local sensitivity be computed at any arbitrary distance $k$ from $T$.

\begin{definition}[Local Sensitivity at Distance $k$ \cite{nissim2007smooth}]
Given a table $T$, the \textit{local sensitivity} of function $f$ is:
\[LS_f(T)^k = \max_{T':\, d(T, T') \leq k} \left\| f(T) - f(T') \right\|_1\]
where $\left\| \cdot \right\|_1$ is the $L_1$ norm.
\end{definition}
A safe approximate upper bound of $LS_f(T)$, $S\_LS_f(T)$, which is insensitive to small variations of data can be obtained by the \textit{smooth sensitivity framework} \cite{nissim2007smooth}.

\begin{definition}[Smooth Sensitivity Framework \cite{nissim2007smooth}]

\begin{equation*}
    \begin{array}{ll}
         S\_LS_f(T) = max_{k=0,1,...n}\{ exp(-\beta k) LS_f(T)^k\}    
    \end{array}
\end{equation*}
where $\beta = \frac{\epsilon}{2log(2/\delta)}$. 


\end{definition}
After a number of $n$ iterations, this upper bound can be used to calibrate noise for the Laplace mechanism to ensure $(\epsilon, \delta)$-DP.\medskip 

\noindent\textbf{DP Properties.} Combining several DP mechanisms is possible, and the privacy accounting is managed using the sequential and the parallel composition properties of DP. 
Let $M_1, \ldots, M_n$ be mechanisms satisfying $(\epsilon_1,\delta_1), \ldots, (\epsilon_n,\delta_n)$ -DP.

\begin{theorem}[Sequential Composition \cite{dp}]\label{th:parallel}

Applying sequentially $M_1, \ldots, M_n$ satisfies $\left( \sum_{j = 1}^{n} \epsilon_j, \sum_{j = 1}^{n} \delta_j \right)$-DP.
\end{theorem}

\begin{theorem}[Parallel Composition \cite{dp}]\label{th:seq}
A mechanism that applies $M_1, \ldots, M_n$ on disjoint parts of the data satisfies:\\
$\left(max_{i \in n}(\epsilon_i),max_{i \in n}(\delta_i)\right)$-DP

\end{theorem}

The \emph{post-processing} property states that it is safe to execute any function on the output of a DP mechanism.
\begin{theorem}[Post-Processing \cite{dp}]\label{th:post}
For any $(\epsilon,\delta)$-DP mechanism $M$ and any function $f$, $f(M)$ satisfies $(\epsilon,\delta)$-DP.
\end{theorem}

In the context of online query answering, each query consumes $(\epsilon,\delta)$ to secure the results. In order to manage/limit the information released to the analyst, a total budget $(\xi,\psi)$ is given which will be consumed by $N$ queries such that $\xi = N\epsilon$ and
$\psi = N\delta$. The analyst can continue sending queries until their total budget is consumed.\medskip

\noindent\textbf{Secure Multiparty Computation (SMC).} it refers to cryptographic protocols that enable a set of independent parties to collaboratively evaluate a query without revealing their private inputs to each other. It also allows them to avoid trusting a third party with the union of their data for query evaluation. However, this safety assurance comes at the cost of resources and processing time. Using SMC is several times slower than insecure alternatives.

\section{Problem Statement}\label{sec:problem}

Given a federated system in which $n$ data providers pool their private data for analysis querying.
Consider a private table $T$ (as in Figure \ref{fig:data_model}) which is horizontally partitioned among data providers as tables $T_1$, $\ldots$, $T_n$.
Each data provider wants to keep the individual tuples of their local table confidential and only the schema of $T$ is public. Suppose an end user sends  the following range query $Q$:

\verb+SELECT COUNT(*) FROM Table WHERE 20 <= Age <= 40+

where $Q$ is performed on the union of tables stored at the data providers, $\cup^n_{i=1} T_i$. However, even though $Q$ may seem very simple at first glance, the big data associated with $T_i$ makes $Q$ very complex and time-consuming. 

To solve the problem of slow query response time, we can resort to Approximate Query Processing (AQP) to find a trade-off between accuracy and speed of results via approximation. One very straightforward technique of AQP is to perform \emph{random sampling}, given a sampling rate $sr$, to obtain a set of tuples from $T$. For example, an end user can request an answer for $Q$ based only on $sr=20\%$ of the entire $T$. Even for a single table $T$, to obtain a good approximation of $Q$, the sampled tuples must contain meaningful data in the ranges of $Q$. Random sampling can be done at the row or cluster level. Although cluster-level sampling is faster than row-level sampling, both have linear performance with respect to sampling rate. The larger the sample, the more accurate and slower the result, and vice versa. 

Consider  $T$ is stored as a set of clusters. To get an accurate estimate of $Q$ when processing a few parts of the data, we use a statistical estimator \cite{book}. To do this, we need to consider the distribution of rows between all clusters. It should be noted that the assumption of a uniform distribution of rows among all clusters is rarely valid in real databases. Indeed, the rows generally follow a skewed distribution. In contrast, unequal probability cluster sampling is more effective at providing better estimates, where the probability of a cluster being sampled is based on the data distribution for $Q$.

Assume that each partition $T_i$ of $T$ is stored using clusters. How to apply the unequal probability cluster sampling in our federated context? Note that each cluster within each data provider should have a specific probability $p$ of being sampled to estimate $Q$, taking into account all other clusters (even those from other data providers). As a result, capturing the inter/intra data distribution will bias the sampling toward clusters or data providers that hold most of the data related to $Q$. We refer to this sampling as global sampling.




The other solution is local sampling, where each data provider computes the sampling probabilities for its clusters (without considering other data providers). In this sampling, the sample size is distributed uniformly on data providers, so it does not require a collaboration between data providers. This lack of global data distribution awareness makes this solution less appealing than global sampling.

To apply global data distribution-aware sampling and approximation, data providers must provide appropriate information about their data to quickly and accurately estimate $Q$. The optimal solution to capture the data distribution in this context is achieved if data providers have access to each other's data and sampling probabilities are computed collectively. This collaboration will lead to an overhead in processing time. The challenge is then to define the summarized and small pieces of information that data providers can share and be sufficient to capture the data distribution while producing negligible overhead. Once this global data distribution is captured, each data provider can locally sample clusters, estimate the query, and send its result. All results from data providers will be added together and the final result will be returned to the end user.

Another dimension of our problem concerns privacy and data protection. In the federated context, the end-to-end privacy property must be guaranteed. This essentially ensures that data is protected (i) during and after query execution, (ii) for intermediate results during collaboration, and (iii) for the final response. Differential Privacy (DP) is a widely accepted privacy model, typically applied to query results to prevent any inference about the presence or absence of individuals. As for the intermediate results produced during collaboration between data providers, they must also be protected, with each data provider seeking to prevent any leakage of information on its table. Even if the exchange is limited to summarized (aggregated) information, there will be no privacy guarantee. Thus, DP can also be used to publish intermediate results between data providers.

An alternative solution to DP is the use of Secure Multiparty Computation (SMC) to implement collaboration between data providers. This solution has two major drawbacks: If data providers use the summary information for sampling in SMC, query approximation (which includes running the query on each cluster) must also be done in SMC because the sampling is based on sensitive information and its results may disclose information to other data providers. Second, SMC relies heavily on cryptography, which will significantly reduce the utility of the query in terms of processing time, thereby diluting the purpose of approximations.

In this work, we aim to provide fast and accurate responses to range queries in a federated setup while preserving end-to-end privacy. The challenges we address are: defining a lightweight sampling algorithm considering data distribution for query approximation in a federated environment and carefully applying Differential Privacy to ensure end-to-end privacy with minimal loss of query accuracy.\vspace{-0.5cm}

\section{Our solution}\label{sec:solution}
\subsection{Overview}\label{sec:overview}
In our proposal, we combine DP with lightweight SMC to protect intermediate results when collaborating between data providers. This allows us to obtain significantly better performance in terms of speed-up and achieve end-to-end privacy, while maintaining high utility answers for online range queries. To achieve these goals, we propose an efficient and lightweight collaboration method, allowing data providers to decide how many samples to extract from each, guided by the summary information shared during this collaboration. To integrate knowledge of the data distribution into our sampling and approximation steps, we use the \emph{probability proportional to size} (pps)  method \cite{book}. Here, the probability $p$ of including (or sampling) a cluster $C$ is determined by the proportion $R$ of rows in $C$ falling within the ranges of the query $Q$. Computing $R$ is expensive and requires similar overhead as running the query. To minimize the processing time of $Q$, we will approximate each $R$ of any cluster $C$ using lightweight \textit{metadata} associated with $C$.

\begin{figure}[h!]
    \centering
    \includegraphics[width=1\linewidth]{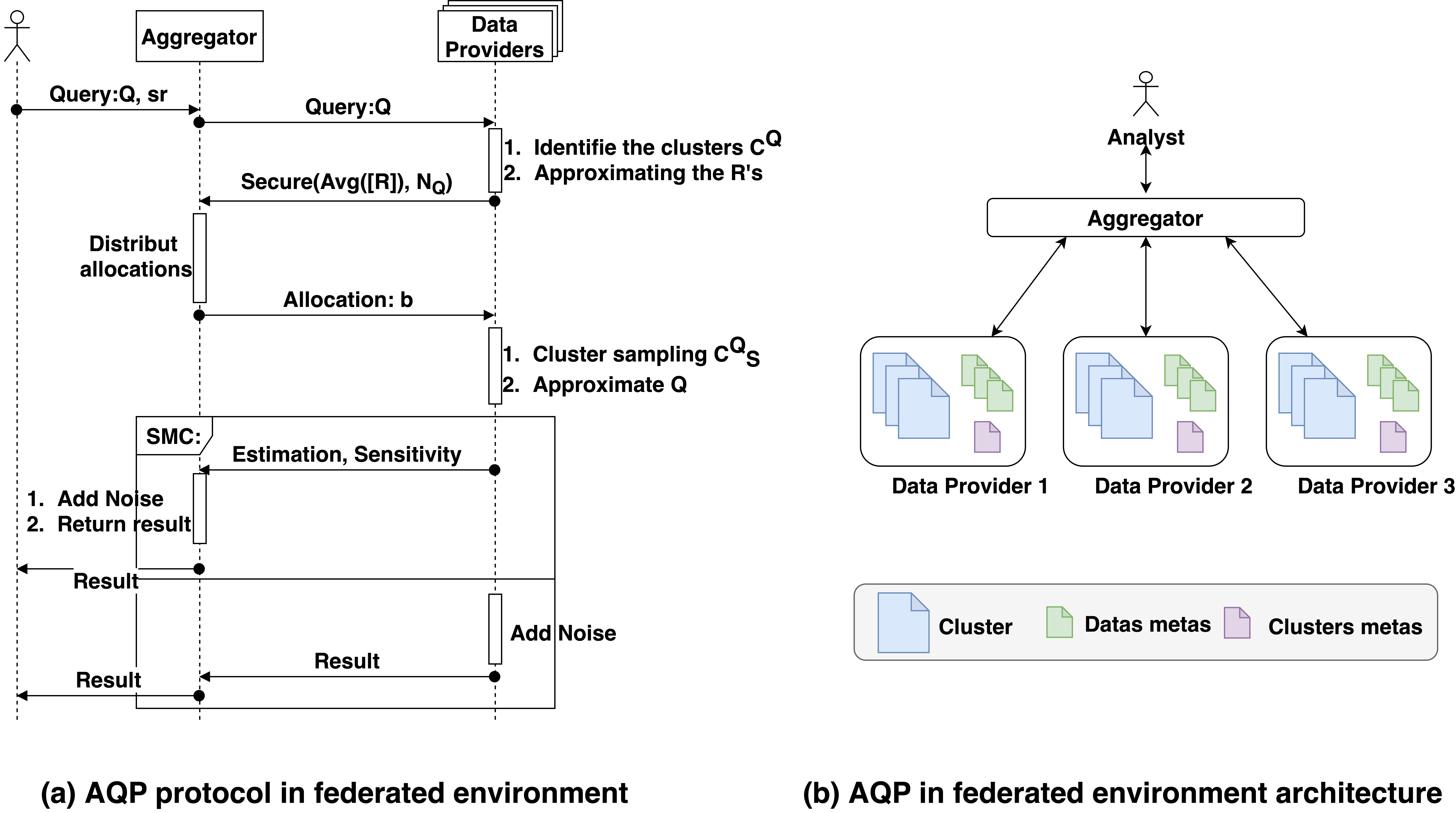}
    \caption{Protocol and Architecture}
    \label{fig:proto}
\end{figure}

Our solution has two main phases: offline data preprocessing and online query answering.
In the offline data preprocessing phase, each data provider constructs global and individual metadata for its clusters. This metadata makes query approximation easier without imposing a significant overhead in terms of processing time. All data providers agree on the same maximum cluster size $S$ (more details are given in Section \ref{sec:discussion}) before initiating the system.
The size $S$ may not reflect the actual size of their clusters, but it would be used to calculate the $R$ of each cluster. The offline phase and metadata creation are detailed in Section \ref{sec:approx}, and Figure \ref{fig:proto} (b) shows the general architecture of our system with each data provider as well as its metadata.

Once preprocessing is complete for all data providers, the system goes online. In the online query response phase, the end user interacts with an aggregator by sending their query $Q$ and desired sampling rate $sr$ and receives a secure response in return. The aggregator manages the rest of the exchanges with the data providers. The query lifecycle (see Figure \ref{fig:proto} (a)) as well as the collaboration (exchange of summary data) are described as follows:
\begin{enumerate}
\item First, the aggregator sends the query $Q$ to the data providers. Each data provider performs two tasks: i) identify the set of clusters $C^Q$ covering $Q$ such that $N^Q = |C^Q|$, ii) compute the proportion $R$ of rows for each $C \in C^Q$. The data provider uses previously stored metadata to avoid overhead when performing these two tasks.
    
\item Each data provider securely (using DP) sends to the aggregator the summarized data needed for collaboration. The number of clusters $N^Q$ and average of proportions $Avg(\widehat{R})$ where $\widehat{R}= \{R_1,...,R_{N^Q}\}$.

\item The aggregator computes and sends the best allocation (sample size $s$) for each data provider while respecting the total sample size given by $sr$.

\item Each data provider tests the condition $N^Q < N^{min}$ in order to compute $Q$ ``regularly'' without approximation. The $N^{min}$ is a threshold set by each data provider to trigger the approximation only if the query is significantly large (more details about $N^{min}$ are given in Section \ref{sec:approx}).

\item If the previous condition does not hold, each data provider randomly and securely with DP samples $C^Q_S$, where $C^Q_S \subset C^Q$.
\item After sampling, each data provider estimates $Q$ over $C^Q_S$ locally and then securely sends the result to the aggregator with DP guarantees.
\item Alternatively, data providers may use SMC to share their local estimations and \textit{"sensitivities"}. Then, the aggregator obliviously sums the estimations and applies DP using the maximum sensitivity before safely releasing the final result.
\end{enumerate}

In Section \ref{sec:approx}, we will focus on the approximation via cluster sampling and the metadata created offline. Afterward, section \ref{sec:fed} will be dedicated to the second phase of our solution. In Section \ref{sec:allo}, we will describe the allocation step and how it preserves the same semantics as the naive (sharing all data) method of collaboration by keeping the sampling data distribution aware without an overhead. In Section \ref{sec:sampling}, we will present the privacy-preserving sampling used by each data provider locally to create $C^Q_S$. In Section \ref{sec:appro_dp}, we detail how to obtain a calibrated DP noise for the end result obtained by using a statistical estimator. Finally in section \ref{sec:accounting}, we explain how the privacy budget for each query is managed and consumed. 

\subsection{Query Approximation and sampling}\label{sec:approx}
As previously mentioned in Section \ref{sec:overview}, our unequal probability sampling is based on the proportion $R$ of rows in cluster $C$ that corresponds to $Q$. Computing the exact $R$ for each cluster is as costly as evaluating the query itself, rendering the approximation useless. Inspired by \cite{sapprox}, we will only approximate $R$ to avoid an overhead in response time.
Given a query $Q$ defined by a set of ranges: $Q = \{\forall d \in D^Q \text{ | } r_d = [l_b^d, u_b^d] \}$ on each dimension, we assume that the dimensions are not correlated (independent). We will compute the sub-proportions $R^d$ on each dimension as follows:
$$R^d = R^{d\ge}(l_b^d)  -  R^{d\ge}(u_b^d)$$
$$ \text{ where :  } R^{d\ge}(x) = \frac{|rows^d \ge x|}{S} \text{ and } S \text{ is the cluster size}$$
The proportion $R^d$ is computed based on the proportions $R^{d\ge}(l_b^d)$ and $R^{d\ge}(u_b^d)$ of records whose dimension $d$ values are $\ge l_b^d$ and  $\ge u_b^d$, respectively. Based on the assumption of independence between dimensions, $R$ can be obtained as follows:
\begin{equation}\label{eq:est_R_p}
R = \prod^{d\in D^Q} R^d \textbf{ and } p_j = \frac{R_j}{\sum_{i=0}^{N^Q} R_i}
\end{equation}

where $N^Q$ is the number of clusters covering $Q$.
The approximated $R$ can then be used to obtain the sampling probabilities $p_j$ for the $jth$ cluster as shown in Equation \ref{eq:est_R_p}. Even this approximation requires a lot of calculations, which may cause similar overhead as the exact $R$. To bypass this limitation, we associate each cluster with a set of metadata that accelerates these computations for any given query (see Algorithm \ref{alg:metas}).

\begin{algorithm}
\caption{Cluster metadata }\label{alg:metas}
\begin{algorithmic}[1]
\Require $T= \{C_1,C_2,...,C_N\}$: Set of clusters
\State $\textbf{Clusters\_metas} \gets create\_global\_meta()$
\For{$\text{each }C \in T$}
    \State $cluster\_meta \gets []$
    \State $\textbf{datas\_meta} \gets create\_datas\_meta\_(C)$
    \For{$\text{each }d \in D$}
        \For{$\text{each }v \in \vert d \vert_C$}
            \State $R^{d\ge}(v) \gets portions\_greater\_(C,d,v)$
            \State $\textbf{datas\_metas}.add(d,v,R^{d\ge}(v))$
        \EndFor
        \State $v^d_{min},v^d_{max} \gets min\_max(\vert d \vert_C)$
        \State $cluster\_meta.add(v^d_{min},v^d_{max})$
    \EndFor
    \State $\textbf{Clusters\_metas}.add(cluster\_meta)$
    \State $save(\textbf{datas\_metas})$
\EndFor
\State $save(\textbf{Clusters\_metas})$

\end{algorithmic}
\end{algorithm}
For each cluster $C$ and for each distinct value $v$ of dimension $d \in D$ in $C \in T$ (Lines 5,6 Algorithm \ref{alg:metas}), $R^{d\ge}(v)$ is stored in the dedicated meta file for the cluster where the entry is in the form $\{ d,v,R^{d\ge}(v) \}$ (Line 8 Algorithm \ref{alg:metas}). These metadata will be used by each data provider to quickly access precomputed proportions that correspond to the range of a given $Q$. Thus significantly reducing the overhead in the online phase. To further improve the performances, Algorithm \ref{alg:metas} stores additional global metadata about the clusters \textbf{Clusters\_metas}, enabling the system to easily identify the clusters $C^Q$ that correspond to $Q$ before even computing the proportions. In a dedicated global file \textbf{\textit{Clusters metas}}, for each dimension $d \in D$ in cluster $C$, Algorithm \ref{alg:metas} (Line 11,13) stores $v^d_{\min}$ ($v^d_{\max}$), the minimum (maximum) value of $d$ in $C$. Based on these metadata in \textbf{\textit{Clusters metas}}, the system is able to focus only on a small subset of the database $C^Q$  that actually contains rows matching $Q$ instead of $T$, thus reducing the processing time of $Q$. The set $C^Q$ is defined as follows:
\begin{equation}\label{eq:C^Q}
    \begin{array}{ll}
          C^Q = \{\forall C \in T  \text{ | } \forall d \in D^Q \text{ , }  [v^d_{min},v^d_{max}] \cap r_d \neq \emptyset  \} &  \\
        \text{ where }  r_d \text{ is the interval of $Q$ in dimension d.}
    \end{array}
\end{equation}

Since we are able to identify the clusters $C^Q$ concerned by $Q$, it only makes sense to approximate $Q$ only when $N^Q$ is bigger than a certain threshold $N^{min}$. This threshold can be set independently by each data provider based on the size of the clusters, the processing time required for a single cluster, and the hardware and software infrastructure.
The cost of saving these metadata is very negligible compared to the actual table and clusters. We used the same data structure like \cite{sapprox} which is very efficient. In Section \ref{sec:eval} we show the space needed for each database.

Once the sampling is applied according to the probability computed using Equation \ref{eq:est_R_p}, the \textit{Hansen-Hurwitz} estimator \cite{book} is used to obtain the final estimation of $Q$. The estimation is done as follows:
\begin{equation}\label{eq:est_p}
    \begin{array}{ll}
         E(Q,C^Q_S) =  \frac{1}{N_S} \sum_{i=1}^{N_S}\frac{Q(C_i)}{p_i}\\
    \end{array}
\end{equation}
where $p_i$ is the sampling probability of the  $ith$ cluster and
$Q(C_i)$ is the query execution result on the $ith$ cluster

\subsection{Federated protocol}\label{sec:fed}
In this section, we will review all the steps of online query approximation and how we were able to carefully integrate DP into each step.

\subsubsection{Allocation phase}\label{sec:allo}
In this step, the data providers $\mathbb{S}$  need to jointly decide the number of clusters to be sampled from each one of them based on the distribution ($R$'s) of data related to $Q$. So upon receiving the query, each data provider identifies $C^Q$ and computes the $R$ for each $C \in C^Q$ using the metadata stored locally. Then each one sends to the $Aggregator$ its $N^Q$ and $\text{Avg}({\widehat{R}})$, where $\widehat{R}$ is the set of $R$'s of the clusters in $C^Q$ and $Avg$ stands for \textit{Average}. $N^Q$ indicates the number of clusters within that data provider that overlap with $Q$, while $\text{Avg}({\widehat{R}})$ shows the average proportion of rows within those clusters that corresponds to $Q$. Based on this information, we obtain an aggregated (summary) view of the data distribution of records corresponding to $Q$ in each data provider. Using these insights, the $Aggregator$ finds the best sample size $s_i$ for $ith$ data provider  using an optimization problem given in Equation \ref{eq:opti_allo} that aims to assign a bigger allocation to the data provider with the most data related to $Q$.

\begin{equation}\label{eq:opti_allo}
    \begin{array}{ll}
        \mbox{ {\bf maximize}} & \sum_{i = 0}^{|\mathbb{S}|} \text{Avg}({\widehat{R}} )_i\times s_i  \\
        \mbox{ {\bf where}} & \sum_{i = 0}^{|\mathbb{S}|} s_i = sr \times \sum_{i = 0}^{|\mathbb{S}|}{N^Q_i}\\ \mbox{ {\bf and }} & sr \in ]0,1[ \text{ is the sampling rate} \\
        \mbox{ {\bf and }} & s_i \in ]1,N^Q_i[ 
    \end{array}
\end{equation}

In Equation \ref{eq:opti_allo}, the data provider that holds the most data related to $Q$ (has the bigger $\text{Avg}({\widehat{R}})_i$) gets more allocation, thus sampling more clusters to approximate $Q$ locally.  This reflects the same behaviour as the original collaboration method (described in Section \ref{sec:problem}): sampling probabilities are computed globally and 
the clusters of the data provider with the bigger $R's$ are more likely to be sampled than others (higher probabilities, Equation \ref{eq:est_p}). So with our collaboration method, we are able to reproduce similar results and behaviour. It is important to highlight that comparing the $\text{Avg}({\widehat{R}})$ from each data provider is only possible because we imposed they use the same $S$ in order to compute the proportions during the metadata creation phase. 

To solve the problem in Equation \ref{eq:opti_allo}, each data provider shares the $N^Q$ and $\text{Avg}({\widehat{R}})$. Both are sensitive pieces of information that may reveal insights about the individuals within the database. Even if the optimisation in Equation \ref{eq:opti_allo} is done over encrypted data, the released allocation $s_i$ might give a data provider insights about the other data providers. To secure the release of this information, each data provider uses \textit{Laplace mechanism} to ensure formal guarantees of privacy. Given a privacy budget of $\epsilon^O$, each data provider perturbs these two values as follows: 
\begin{equation}
    \begin{array}{ll}
         \widetilde{\text{Avg}}({\widehat{R}}) = \text{Avg}({\widehat{R})} + \text{Lap}(\frac{\Delta_{\text{Avg}(\widehat{R}})}{\epsilon^O / 2})  \\
         \widetilde{N^Q} = N^Q + \text{Lap}(\frac{1}{\epsilon^O / 2}) 
    \end{array}
\end{equation}

where the sensitivity of $N^Q$ to the absence/presence of an individual is 1, and the sensitivity of $\text{Avg}(\widehat{R})$, is $\Delta_{\text{Avg}(\widehat{R}} $.

\begin{theorem}[Sensitivity of estimator $\Delta_{\text{Avg}(\widehat{R}}$]\label{th:unbounded_est}
For any two neighbouring databases $T$, $T'$ the sensitivity of  $\text{Avg}({\widehat{R})}$ is defined as: 
$$\Delta _{Avg(\widehat{R})} = max(\frac{\Delta_R}{N^{min}},\frac{1}{N^{min} + 1}) \\
\text{ where: }\Delta_R = 1 - (1 - \frac{1}{S})^{\left| D \right|}$$

\end{theorem}

 The proof is given in appendix \ref{apn:collab_data}.
 
 With this perturbation, the collaboration between data providers for deciding the allocation does not reveal any sensitive information. So the optimization problem is formulated as follows:

\begin{equation}\label{eq:opti_allo_dp}
    \begin{array}{ll}
        \mbox{ {\bf maximize}} & \sum_{i = 0}^{|\mathbb{S}|} \widetilde{\text{Avg}}({\widehat{R_i}}) \times s_i  \\
        \mbox{ {\bf where}} & \sum_{i = 0}^{|\mathbb{S}|} s_i = sr \times \sum_{i = 0}^{|\mathbb{S}|}\widetilde{N^Q_i}\\ \mbox{ {\bf and }} & sr \in ]0,1[ \text{ is the sampling rate} \\
        \mbox{ {\bf and }} & s_i \in ]1,\widetilde{N^Q_i}[ 
    \end{array}   
\end{equation}

The test of $N^Q < N^{min}$ comes after the allocation (collaboration) phase in order to encourage all data providers to participate. Otherwise, if a data provider does not participate in allocation because locally approximating $Q$ is not possible, this may reveal information about the size of its data to other data providers.

\subsubsection{Sampling phase}\label{sec:sampling}

After the allocation phase, each data provider receives an allocation $s$: the number of clusters to process for the $Q$ approximation. Using the $\widehat{R}$ computed locally, the data provider computes the sampling probabilities for $C^Q$ and then performs unequal probability sampling to randomly select $s$ clusters. Since the sampling probabilities are computed based on the rows (individuals) in the database, the result of the sampling (choices) may leak information about the presence/absence of any individual. To guarantee DP, our system uses the Exponential Mechanism (EM) to select the $s$ clusters $C^Q_S \subset C^Q$ (Algorithm \ref{alg:dp_sample}) while consuming $\epsilon^{S}$ privacy budget.

\begin{algorithm}
\caption{$EM\_sampling$}\label{alg:dp_sample}
\begin{algorithmic}[1]
\Require $C^Q: \text{set of clusters}$, $\widehat{R}: \text{set of corresponding $R's$ to $C^Q$}$, $s: \text{sample size}$, $\epsilon^{S}: \text{total budget}$
\State $P \gets \text{get\_sampling\_probabilities}(\widehat{R})$ \Comment{ Equation \ref{eq:est_R_p}}
\State $P^{EM} \gets []$
\State $\epsilon^s \gets \epsilon^{S} / s$
\For{$i \in [1, N^Q]$}
  \State $P^{EM}[i] \gets \text{exp}\left(\frac{\epsilon^s \times P[i]}{2 \times \Delta p}\right)$
\EndFor
\State $C^Q_S \gets \text{random\_choice}(C^Q, P^{EM}, s)$
\State $\textbf{Return } C^Q_S, P$
\end{algorithmic}
\end{algorithm}

The score of the $ith$ cluster $C_i \in C^Q$ is its own sampling probability $p_i$ (Algorithm \ref{alg:dp_sample} line 1),  which means the scoring function $L$ of EM is defined by the computation in Equation \ref{eq:est_R_p}. So to calibrate the noise (randomness) of EM, we must find the sensitivity of this function $L$ to the absence/presence of any individual in the database.

Consider two neighbouring databases $T$ and $T'$, where $T'$ is obtained by adding any random record (which represents an individual) to $T$ at any possible cluster. Given a range query $Q$, in order to measure $\Delta p_i$ (sensitivity of $p_i$, which is the same as $L$) we assume the worst case scenario for $T$ and $T'$: all clusters of $C^Q (C^Q \subset T$) each have a record that corresponds to $Q$. In this case, their probabilities are the same: $p  = \frac{1}{N^Q}$. In $T'$, one record is added to another cluster $C'$ outside of $C^Q$ that matches $Q$. Thus $C'^Q =C^Q \cup \{C'\}$ and $N'^Q = N^Q + 1$, and for $Q$ all the clusters have the same sampling probability: $p' = \frac{1}{N^Q + 1}$. So the $\Delta p$ can be computed as follows:

\begin{equation}
    \begin{array}{ll}
         \Delta p \leq \left|\frac{1}{N^Q} - \frac{1}{N^Q +1}\right| \implies 
\Delta p \leq \frac{1}{N^Q \times (N^Q+1)}
    \end{array}
\end{equation}

We notice that $\Delta p$ is dependent on the query $Q$. To find the global maximum value for $\Delta p$, we replace the $N^Q$ by its minimum possible value $N^{min}$.

\begin{theorem}[Sensitivity of sampling probability]\label{th:sens_p}
For any two neighbouring databases $T$, $T'$ the sensitivity of the sampling probability of any cluster $C$ is bounded by : 
$$\Delta p = \max_{T,T'} \left\| p_C(T) - p_C(T') \right\|_1 = \frac{1}{N^{min} \times (N^{min}+1)}$$
where $\left\| . \right\|_1$ is the $L_1$ norm.
\end{theorem}

In Algorithm \ref{alg:dp_sample} line 5, this sensitivity $\Delta p$ is used for sampling using $EM$. To manage the total budget $\epsilon^{S}$ allocated for $EM$ in order to safely make $s$ selections (Algorithm \ref{alg:dp_sample} line 7), we set $\epsilon^s = \frac{\epsilon^{S}}{s}$ the budget of each random selection (Algorithm \ref{alg:dp_sample} line 3).

\subsubsection{Approximation phase}\label{sec:appro_dp}
To obtain the final result from $C^Q_S$, each data provider uses the estimator $E$ defined in Equation \ref{eq:est_p}. In order to release the final results securely and have DP privacy guarantees, a well-calibrated noise will be added to the final answer using \textit{Laplace Mechanism}.
To apply \textit{Laplace Mechanism}, we need to find the sensitivity $\Delta_{E}$ of the estimator. Let us define $\mathbb{E}(C,Q,p) = \frac{Q(C)}{p}$. We can re-write $E$ as follows :
\begin{equation}
    \begin{array}{ll}
       E(Q,C^Q_S) = \frac{1}{s} \sum_{i=1}^{s} \mathbb{E}(Q,C_i,p_i) \\
    \mbox{{\bf where }}  \text{$s$ is the size of } C^Q_S
    \end{array}
\end{equation}

Which implies that :
\begin{equation}\label{eq:avg_sens_e}
    \Delta_E = \frac{1}{s} \sum_{i=1}^{s} \Delta_{\mathbb{E}}
\end{equation}
To find $\Delta_E$, we will focus on finding $\Delta_{\mathbb{E}}$, and deduce $\Delta_E$ afterwards based on this implication.
Given that $\mathbb{E}(C,Q,p) = \frac{Q(C)}{p}$ is a fraction of two real values, it gives a hint that its sensitivity might be unbounded similarly to $Average$ operator \cite{near_abuah_2021}. Upon further analysis (see appendix \ref{apn_estimator}), we find that $\Delta_{\mathbb{E}}$ is unbounded, which implies $\Delta_E$ is also unbounded.

\begin{theorem}[Sensitivity of estimator $\mathbb{E}$]\label{th:unbounded_est}
For any two neighbouring databases $T$, $T'$ the sensitivity of the estimator $\mathbb{E}$ for any cluster $C$ and query $Q$ is unbounded: 
$$\Delta_{\mathbb{E}} = \max_{T,T'} \left\| \mathbb{E}(Q,C) - \mathbb{E}(Q,C') \right\|_1 \ge \frac{N \times S^D}{2} -1$$
where $\left\| . \right\|_1$ is the $L_1$ norm.
\end{theorem}
See appendix \ref{apn_estimator} for a proof of theorem \ref{th:unbounded_est}.

Given that a global sensitivity does not exist, we resort to the \textit{Local Sensitivity (LS)} which is measured based on the database instance $T$. For any database $T'$ neighbouring to $T$  obtained by adding 1 row (one individual) that matches the query $Q$, we can distinguish four scenarios for a cluster $C \in C^Q$ (we focus on one cluster $C$ because we are looking for $\Delta_{\mathbb{E}}$) that might affect $\mathbb{E}$:
\begin{itemize}
    \item Scenario 1: Cluster $C$ did not receive the new row, but another cluster did.
    \item Scenario 2: Cluster $C$ did receive the new row.
    \item Scenario 3: Cluster $C$ did not receive the new row but another cluster has been added to $C^Q$, such that $N'^Q = N^Q+1$.
    \item Scenario 4: Cluster did receive the new individual, but only add $+1$ to the $Measure$ attribute of existing aggregate row.
\end{itemize}

Our aim is to find the upper bound of $LS_{\mathbb{E}}$, thus we must consider the distance that provides the largest sensitivity.
An analysis of each of these scenarios (see Appendix \ref{apn:decide}) showed that under a certain condition, either scenario 1 or scenario 4 will yield the biggest distance. For a given cluster $C$, we can choose the \textit{Dominant scenario} (which will yield the biggest $LS_{\mathbb{E}}$) between scenarios 1 and 4 without needing to compute any of them.
\begin{theorem}[Dominant distance LS]\label{th:domin_dist}
the neighbouring scenario 1 will give bigger distance than scenario 4 iff: 
$$Q(C) > \frac{\sum^{R \in \widehat{R}}R}{\Delta_R}$$
\end{theorem}
See Appendix \ref{apn:decide} for proof.\\
Since the $LS_{\mathbb{E}}$ is computed based on $T$, it cannot be used directly to inject noise because the scale of the noise may reveal sensitive information about $T$ \cite{near_abuah_2021}. To avoid such information leakage, we will use the \textit{smooth sensitivity framework} \cite{nissim2007smooth} for finding a safer upper bound $S\_LS_\mathbb{E}$ for the \textit{local sensitivity} $LS_\mathbb{E}$. So we redefine our $LS_{\mathbb{E}}$ in terms of a distance $k$ between $T$ and $T'$:
\begin{itemize}
    \item Scenario 1: $  LS^k_{\mathbb{E}} = k \times \frac{Q(C)\times \Delta_R}{R}$
    \item Scenario 4: $LS^k_{\mathbb{E}} = k \times \frac{1}{p}$

\end{itemize}
See Appendix \ref{apn:decide} for proof.

The safe smooth upper $S\_LS_\mathbb{E}$ is defined as follows:
\begin{equation}\label{eq:s_ls}
    \begin{array}{ll}
         S\_LS_\mathbb{E} = max_{k=0,1,...n}\{ e^{-\beta k}\times LS^k_{\mathbb{E}}\}    
    \end{array}
\end{equation}
$\text{ where } \beta = \frac{\epsilon^E}{2 \times\ln(2/\delta)} $ and $(\epsilon^E,\delta)$ is the privacy budget allocated for releasing the final result.

\begin{algorithm}[h]
\caption{$Estimate\_Q$}\label{alg:est_q}
\begin{algorithmic}[1]
\Require $Q:query, C^Q_S:clusters,(\epsilon^{E},\delta):budget, SMC:bool$
\State $result \gets \text{approximate\_Q}(Q,C^Q_S)$ \Comment{ Equation \ref{eq:est_p}}
\State $S\_LS \gets []$
\For{$i \in [1, N^Q_S]$}
  \State $S\_LS[i] \gets smooth\_LS(Q,C^Q_S[i],\epsilon^E,\delta)$ \Comment{ Equation \ref{eq:s_ls}}
\EndFor
\State $LS\_smooth \gets average(S\_LS)$ \Comment{ Equation \ref{eq:avg_sens_e}}
\If {SMC}
    \State $send\_secure(result,LS\_smooth)$
\Else
    \State $dp\_result \gets result + Lap(\frac{2 \times LS\_smooth}{\epsilon^E})$
    \State $send(dp\_result)$
\EndIf
\end{algorithmic}
\end{algorithm}

Based on the definitions we gave for $LS^k_{\mathbb{E}}$, the computational overhead to compute the \textit{smooth sensitivity} for each cluster $C \in C^Q_S$ is very negligible because: i) All the $R$'s and $p$'s are computed before this step, and will be reused for each iteration over $k$; ii) the maximum value of $k$ (steps) is also bounded by $k = \frac{1}{1 - e^{\beta}} +1$ (see Appendix \ref{apn:bound_k} for proof), which guarantees that the process will terminate; iii) Theorem \ref{th:domin_dist} allows to determine which scenario is \textit{dominant} for any given cluster, thus only computing one $S\_LS_\mathbb{E}$.

Algorithm \ref{alg:est_q} describes the process of estimating $Q$ over the subset of cluster $C^Q_S$. It  \ref{alg:est_q}  starts in line 1 by estimating  $Q$ according to Equation \ref{eq:est_p}. Then it proceeds to compute the \textit{smooth sensitivity} (Lines 2-6), where the function $smooth\_LS$ is responsible for computing the smooth sensitivity $S\_LS_\mathbb{E}$  for each cluster $C \in C^Q_S$ as described in Equation \ref{eq:s_ls}. Depending on the chosen setup by the data providers, either they compute and send a DP result to the \textit{aggregator} (Algorithm \ref{alg:est_q}, Lines 10–11) and the \textit{aggregator} returns the sum to the user.
The second option is that data providers share their estimations and computed sensitivities (Algorithm \ref{alg:est_q}, Line 8) with the $Aggregator$ securely using SMC, and obliviously compute the sum of estimations and the max sensitivity to perturb the final result with \textit{Laplace Mechanism}.\vspace{-0.3cm}

\subsection{Privacy accounting}\label{sec:accounting}
In the online query answering settings under DP, the end user is limited by a total privacy budget of $(\xi,\psi)$. For each query $Q$, a budget $(\epsilon,\delta)$ is consumed in order to publish the answer and the end user can interact with system as long as the total budget $(\xi,\psi)$ is not consumed. In this section, we will track the privacy budget $\epsilon$ consumption for each query.

In our proposed protocol the data providers do not share their data, and $Q$ is processed (data access and publishing) in parallel by each data provider. We can just track the consumption on one data provider, and based on the \textit{parallel composition property} of DP we can deduce the budget consumption for $Q$ on the full system. A data provider starts by publishing the  $N^Q$ and $\text{Avg}({\widehat{R}})$ using \textit{Laplace mechanism} for the allocation phase, while consuming a total budget of $\epsilon^O$. Based on the \textit{post-processing property} of DP,
obtaining the sample size $s$ is DP. Afterwards, each data provider uses \textit{Exponential Mechanism} to sample a subset $C^Q_S \subset C^Q$ while consuming a budget of $\epsilon^S$. To publish an estimation of $Q$, each data provider uses \textit{Laplace mechanism} once more, and consumes a budget of $\epsilon^E$.  The final step does not in fact guarantee pure DP, since the smooth sensitivity has a $\delta$ \textit{failure probability}. Based on the \textit{sequential composition property} of DP, 
the total budget  is: $(\epsilon = \epsilon^O + \epsilon^S + \epsilon^E, \delta)$. Given the \textit{parallel composition property}, the budget consumption for $Q$ is $(\epsilon,\delta)$.

In case the data providers used SMC to inject a single noise, based on \textit{parallel composition property} we deduce that data providers consumed $\epsilon^O + \epsilon^S$ for the local computation. Afterwards they collectively consumed (once) $\epsilon^E$ for publishing the result. By the \textit{sequential composition property} of DP, the budget consumption for $Q$ is $(\epsilon = \epsilon^O + \epsilon^S + \epsilon^E, \delta)$.

Based on these results, a set of hyperparameters can be set in our system (by database admin for example) that regulates the $\epsilon$ budget distribution at each step of the query processing.\\
Let $hp_1,hp_2 \text{ and } hp_3$ be this set of hyperparameters
(where $hp_i \in ]0,1[$ and $hp_1 + hp_2 + hp_3 = 1$) such that : $\epsilon^O = hp_1 \times \epsilon $, $\epsilon^S = hp_2 \times \epsilon $ and $\epsilon^E = hp_3 \times \epsilon $.

\section{Evaluation}\label{sec:eval}
\subsection{Setup}
\noindent\textbf{Datasets.} We used two big datasets: 
(i) \emph{Adult} \cite{adult} contains demographic and income information for individuals with $15$ dimensions and $48 \times 10^3$ records, synthetically scaled up $4 \times 10^6$ records. 
(ii) \emph{Amazon Review} \cite{amazon} is about reviews from Amazon clients across different product categories, with only three ``range querable'' dimensions and $231 \times 10^6$ records ($\sim120$ Gb). We synthetically added three randomly populated dimensions and random records to reach $4 \times 231 \times 10^6$ records.

A count tensor with column \emph{Measure} is created from each dataset, aggregating six dimensions of \emph{Adult} and one dimension of \emph{Amazon Review}.

\noindent\textbf{Queries and Workloads.} We generated random ranges for the queries and ran only those that lead to the approximation ($N^{min} < N^Q$) on all data providers. A workload $(m,n)$ is a set of $m$ \emph{distinct} queries with ranges over $n$ dimensions.

\noindent\textbf{Metrics.} An online query is useful if it has a low error rate and low processing time.
To measure the query error, we used $\text{\textit{Relative error}} = \frac{|\text{answer} - \text{estimation}|}{\text{answer}}$. 
For performance in terms of \textit{response time}, we used: $\text{\textit{Speed-UP}} $= $\frac{\text{time of normal computation}}{\text {time of estimate computation}}$.

\noindent\textbf{Configuration.} In our experiments, we assumed that there are one aggregator and four data providers and that each data provider has its own database. 
Datasets \emph{Adult} and \emph{Amazon Review} are horizontally partitioned equally across data providers.

\noindent\textbf{Source code.} Based on PostgreSQL\footnote{\url{https://developers.google.com/optimization}}, our solution\footnote{\url{https://github.com/AlaEddineLaouir/Federated-Range-Queries.git}} coded in Python uses the libraries: (i) OrTools\footnote{\url{https://developers.google.com/optimization}} as solver;
(ii) Pyro5\footnote{\url{https://pyro5.readthedocs.io/en/latest/index.html}} as communication medium; and, 
(iii) MPyC\footnote{\url{https://mpyc.readthedocs.io/en/latest/mpyc.html}} as SMC environment.
Our implementation is a proof-of-concept in which the clusters of the original table are other smaller tables.

\noindent\textbf{Hyperparameters.} In our experiments, the total privacy budget $(\epsilon,\delta)$
for each query is set with $\delta = 10^{-3}$ and $\epsilon = 1$ (unless other values are indicated for $\epsilon$).
The budget $\epsilon$ is shared between each step of our solution as follows: $\epsilon^O = 0.1 \times \epsilon$, $\epsilon^S = 0.1 \times \epsilon$ and $\epsilon ^E = 0.8 \times \epsilon$.
To get clusters of the same size, we set the cluster size $S$ to
$1\%$ and $0.5\%$ of the total size $T^a$ of each data provider for \emph{Adult} and \emph{Amazon Review}, respectively.

\noindent\textbf{Metadata space allocation}. The metadata for \textit{Amazon Review} dataset was about 11 MB (56 KB/cluster). As for \textit{Adult} dataset, it occupied 6.4 MB (64 KB/cluster).

\noindent\textbf{Hardware\footnote{Grid5000: Grisou cluster \url{https://www.grid5000.fr/w/Nancy:Hardware}}.} For each of the data providers and the aggregator, we allocated a dedicated server with the following configuration: $2$ X Intel Xeon E$5-2630$ v$3$ $8$ cores/CPU x$86\_64$, RAM $128$ GB and $1.2$ TB HDD, and a network with $1$ Gbps + $4$ x $10$ Gbps (SR‑IOV).

\subsection{Dimension-based analysis}
In these experiments, we evaluated the impact of the number of dimensions in queries on accuracy.
To this end, we generated random workloads $(m,n)$ with $m =100$ distinct queries (\verb+SUM+ and \verb+COUNT+) and dimension $n\in [2,7]$ for \emph{Adult} and $n \in [2,5]$ for \emph{Amazon Review}.
For the sampling rate, we set it to $5\%$ and $20\%$ for \emph{Amazon Review} and \emph{Adult} datasets, respectively.

\begin{figure}[t]
    \centering
    \includegraphics[width=1\linewidth, height=7cm]{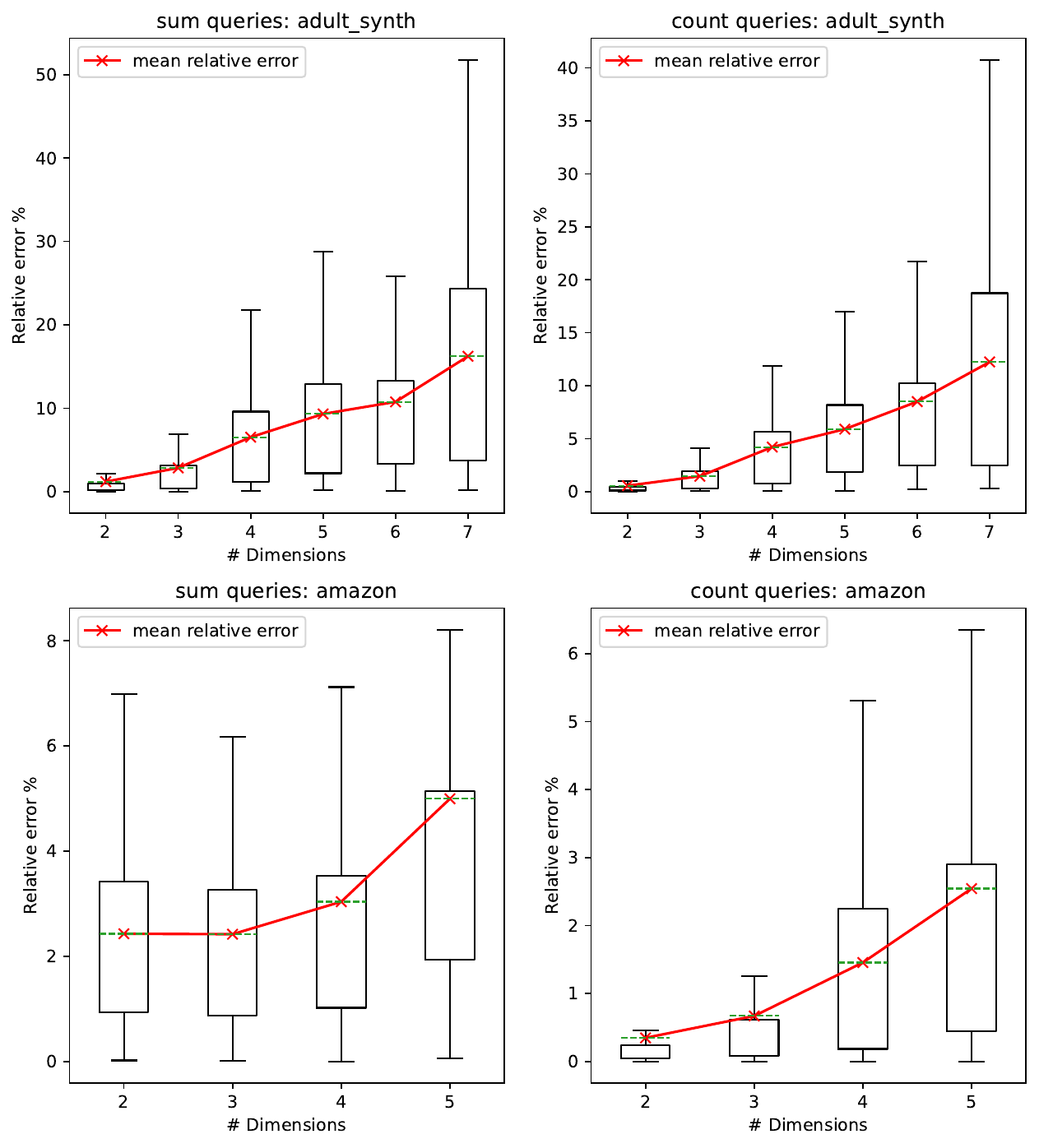}
    
    \caption{Dimension-based analysis}
    \label{fig:dim_an}
\end{figure}

The results presented in Figure \ref{fig:dim_an} show that our solution achieves very high accuracy for \verb+COUNT+ and \verb+SUM+ queries.  The relative error is less than $2.5\%$ (resp. $11\%$) on average for \verb+COUNT+ queries on \emph{Amazon Review} (resp. \emph{Adult}).
As for \verb+SUM+ queries, the error is less than $5\%$ (resp. $17\%$) on \emph{Amazon Review} (resp. \emph{Adult}).
This performance difference is due to the size difference between the databases. In big tables, query results are larger (contain more data), therefore less affected by \textit{Laplace Mechanism} noise.
Interestingly, the results also indicate  that queries become more accurate as the number of dimensions decreases. Specifically, with workloads having only $2$ dimensions on both datasets, we reached an error close to $0\%$.
This observed behavior corresponds to our expectations. Because in Equation \ref{eq:est_R_p}, we approximate $R$ of each cluster and the accuracy of this approximation improves as the number of dimensions decreases, bringing the approximation closer to the exact $R$. Thus, we have more accurate sampling probabilities which affect the estimation of the final result.
For the speedup, the results in Figure \ref{fig:speed_amazon} show that the higher the number of dimensions, the less speedup is gained. From the results in Figure \ref{fig:speed_amazon}, the speedup drops from approximately $8x$ to $6x$ as the number of dimensions increases from $2$ to $5$ on \emph{Amazon Review} dataset. This drop is attributed to the sampling probabilities approximation phase, where our algorithm looks up the preprocessed metadata. The higher the number of dimensions, the more metadata it needs to look up. However, this effect becomes negligible on larger databases. Because even in these results, the speedup remains very significant.

\subsection{Sampling rate-based analysis}
\begin{figure}[h]
    \centering
    \includegraphics[width=1\linewidth, height=7cm]{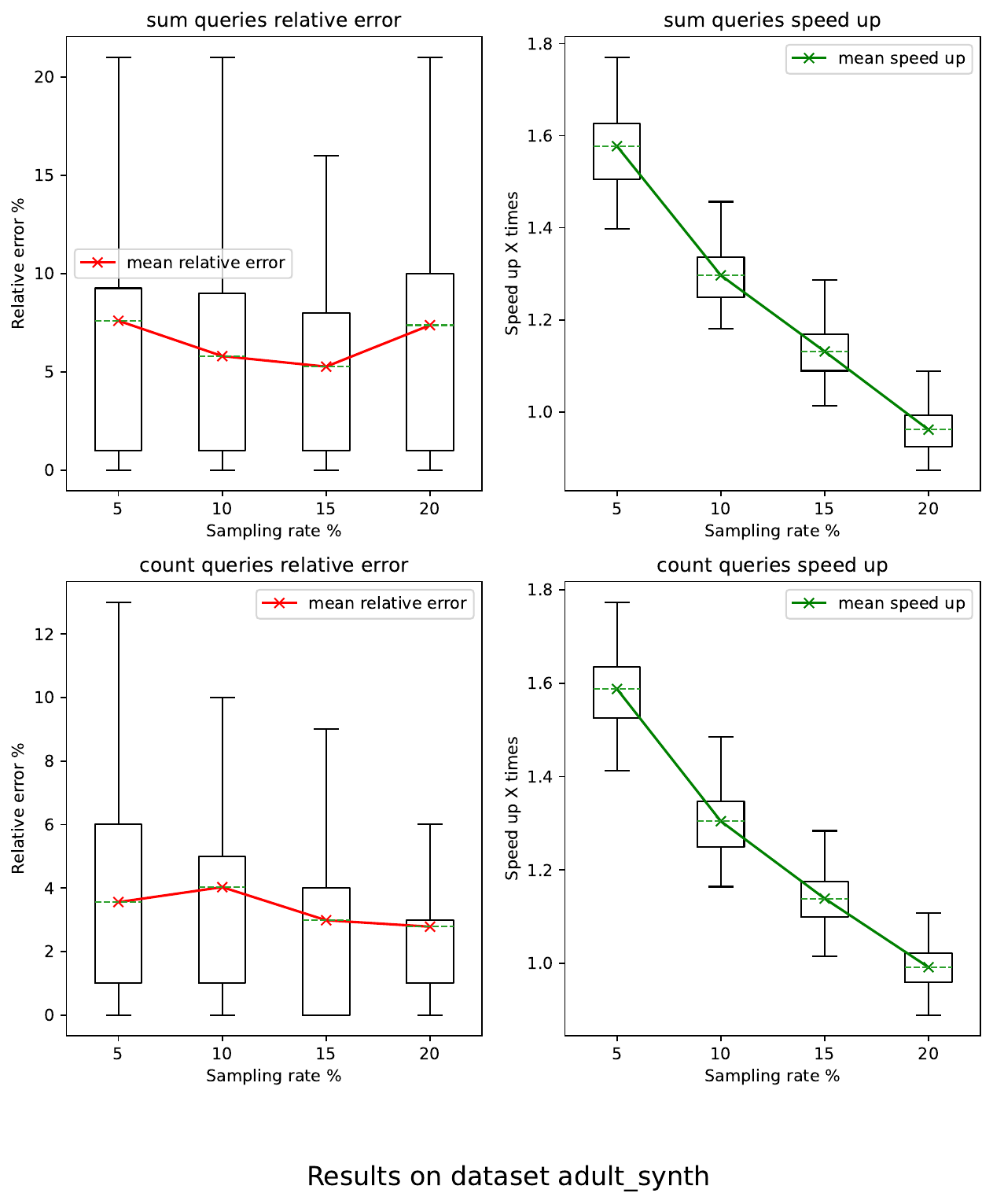}
    \includegraphics[width=1\linewidth, height=7cm]{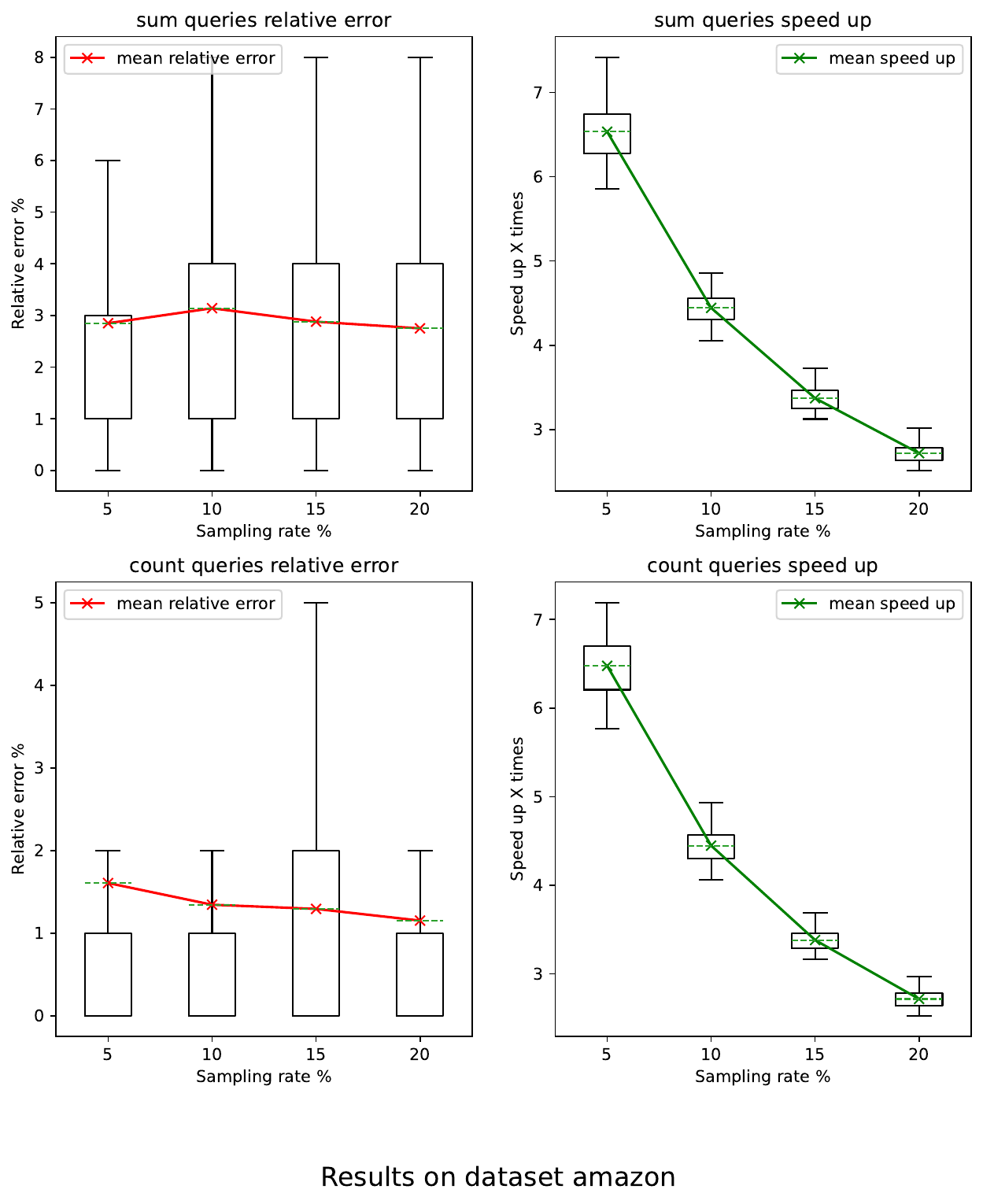}
    \caption{Sampling rate-based analysis}
    \label{fig:sr_an}
\end{figure}

In this analysis, we examined the effect of sampling rate on query quality. For each database, we generated two random workloads for \verb+COUNT+ and \verb+SUM+ queries of $m=100$ and $n =4$.
We varied the sampling rate between $5\%$ and $20\%$ for each experiment and measured the quality obtained in terms of accuracy and speed-up. From the results in Figure \ref{fig:sr_an}, we observe that a higher sampling rate provides slightly better accuracy: reaching a relative error of less than $1\%$ with a $20\%$ sampling rate for \verb+COUNT+ queries on \emph{Amazon Review} dataset.

Regarding the speed-up, we note that our solution reaches up to a $7x$ compared to a normal execution (without approximation) on \emph{Amazon Review} (with $4$ dimensional queries). Additionally, the speed-up gains in \emph{Amazon Review}  are $4x$ more significant than those in \emph{Adult}. 
This result indicates that our solution provides more speed for larger datasets. Also based on the results in Figure \ref{fig:sr_an}, the tradeoff between speed-up and accuracy is noticeable. We observe that the larger the sampling, the less the speed-up is gained. On the other hand, accuracy improves with higher sampling rates. We can say that, based on the results shown in this experiment, accuracy gains with higher sampling are very costly in terms of speed-up. But it is up to the users (data analysts) to define the sampling rate according to their needs.

\subsection{Privacy budget-based analysis}
\begin{figure}
    \centering
    \includegraphics[width=1\linewidth, height=7cm]{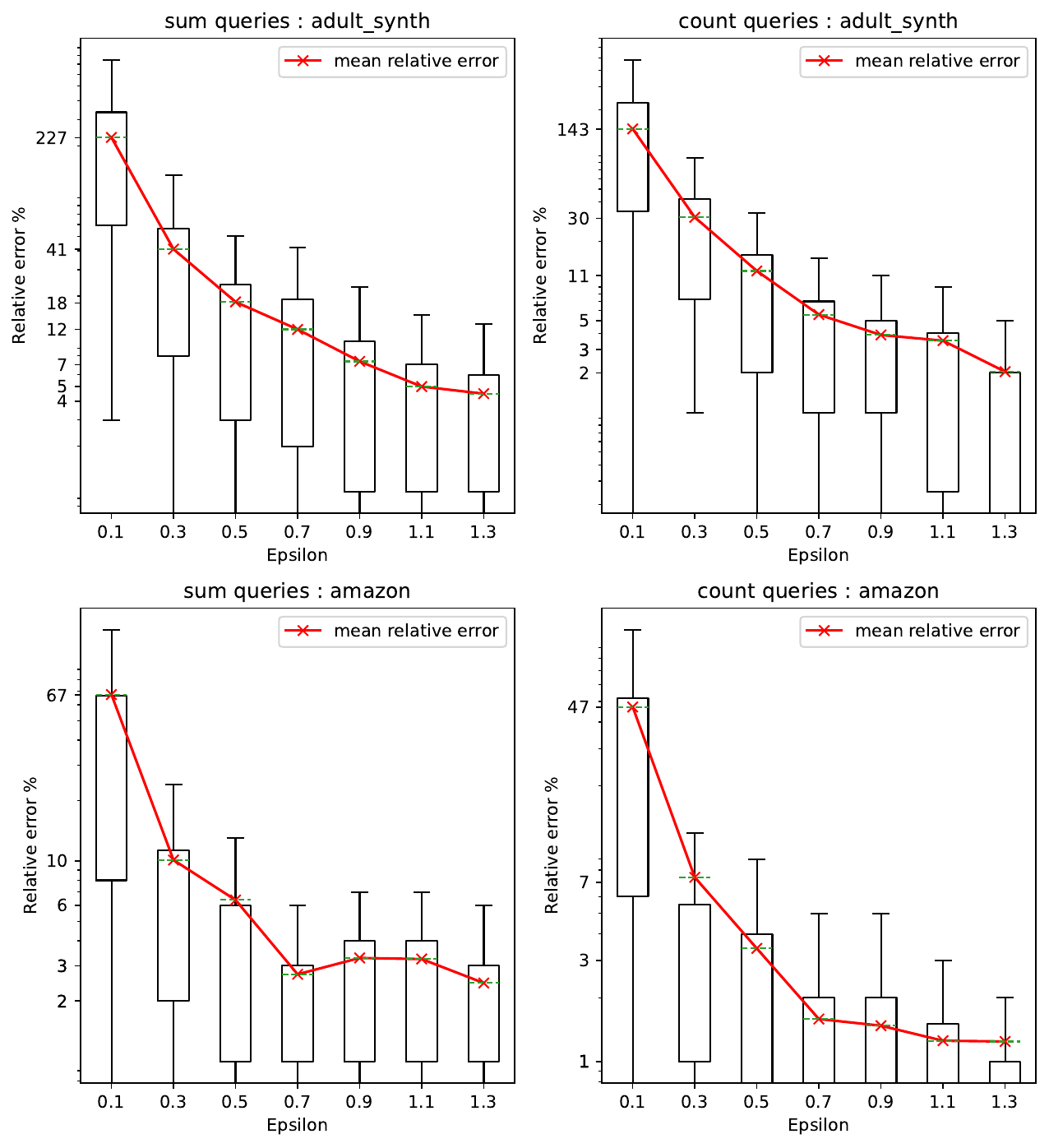}
    \caption{Epsilon-based analysis}
    \label{fig:ep_an}
\end{figure}

In these experiments, we analyzed the effect of the privacy budget $\epsilon$ on query quality. We generated two random workloads of $m=100$ and $n =4$ for \verb+COUNT+ and \verb+SUM+ queries and set the sampling rate to $5\%$ and $10\%$ for \textit{Amazon Review} and \textit{Adult}, respectively.
We varied $\epsilon$ between $0.1$ and $1.3$ and captured the performance on each workload. From the results in Figure \ref{fig:ep_an}, we can immediately observe the typical trend of any DP mechanism (larger $\epsilon$ leads to better accuracy).

\begin{figure}
    \centering
    \includegraphics[width=1\linewidth, height=7cm]{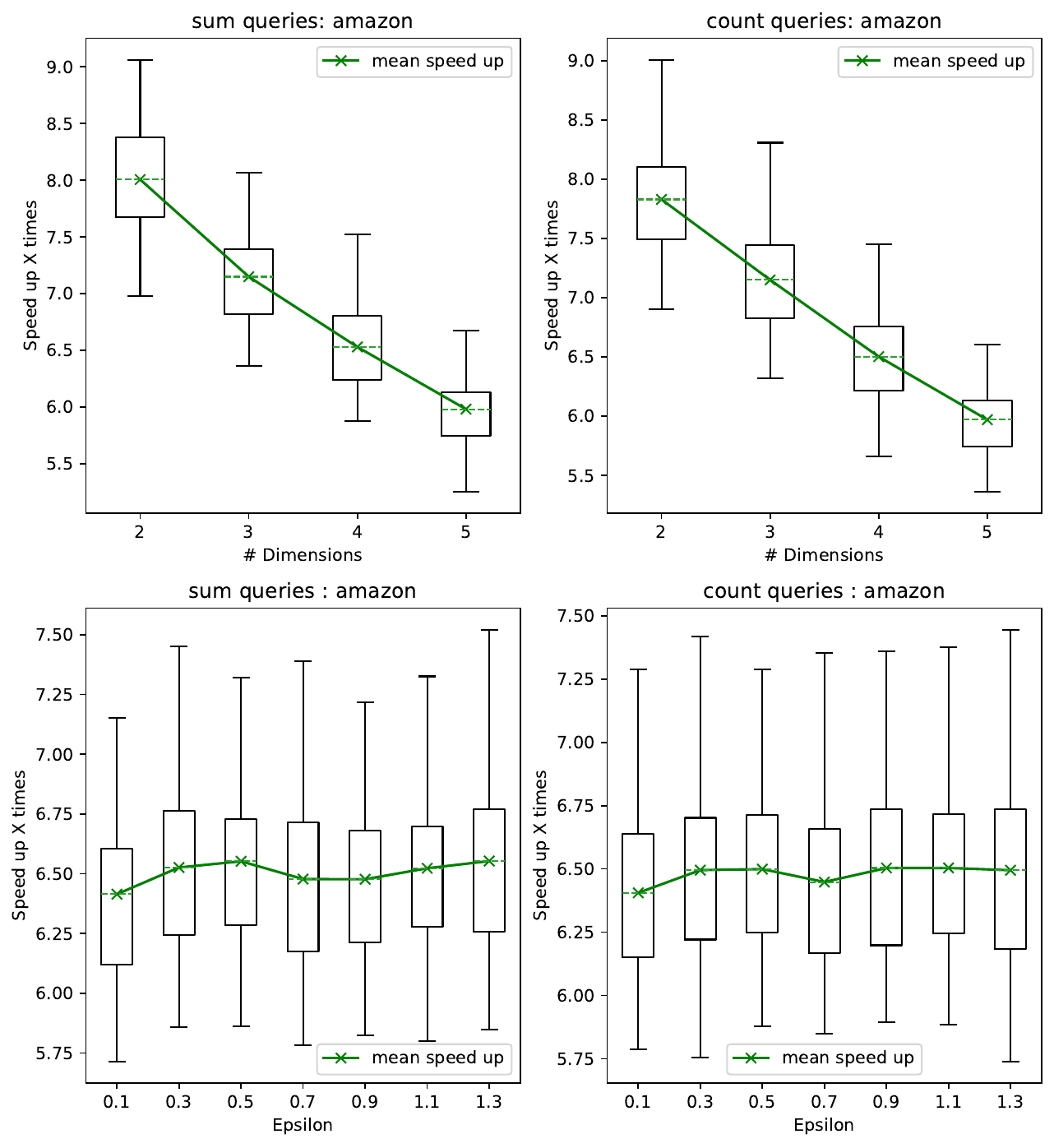}
    \caption{Impact of dimension and $\epsilon$ on speed-up}
    \label{fig:speed_amazon}
\end{figure}

Interestingly, \verb+SUM+ queries are able to provide better utility (lower relative error) than \verb+COUNT+ queries. This happens because \verb+SUM+ queries yield more substantial results (larger query responses) than \verb+COUNT+ queries, making them less affected by noise added to the response.
A similar observation applies when comparing results between the two databases, with workloads on \emph{Amazon Review} preserving more accuracy than those on \emph{Adult}. This is attributed to the fact that the \emph{Amazon Review} dataset is much larger than \emph{Adult}, causing queries to be less affected by the added noise.
Based on this observation, we can predict that as the database size increases, the accuracy of our solution will improve by using smaller values for $\epsilon$. Regarding speed-up, the results in Figure \ref{fig:speed_amazon} show that $\epsilon$ levels have no effect.


\subsection{SMC vs DP in terms of sharing results}
To examine the performance of our SMC-based solution to share final results, we conducted experiments using an \emph{Adult} dataset split across four data providers.
We generated five random two-dimensional \verb+COUNT+ queries. Each query was repeated five times (with and without SMC) and we measured the speed-up and the the range of noise added using the \textit{Laplace mechanism} at each iteration.

\begin{figure}[h]
    \centering
    \includegraphics[width=1\linewidth]{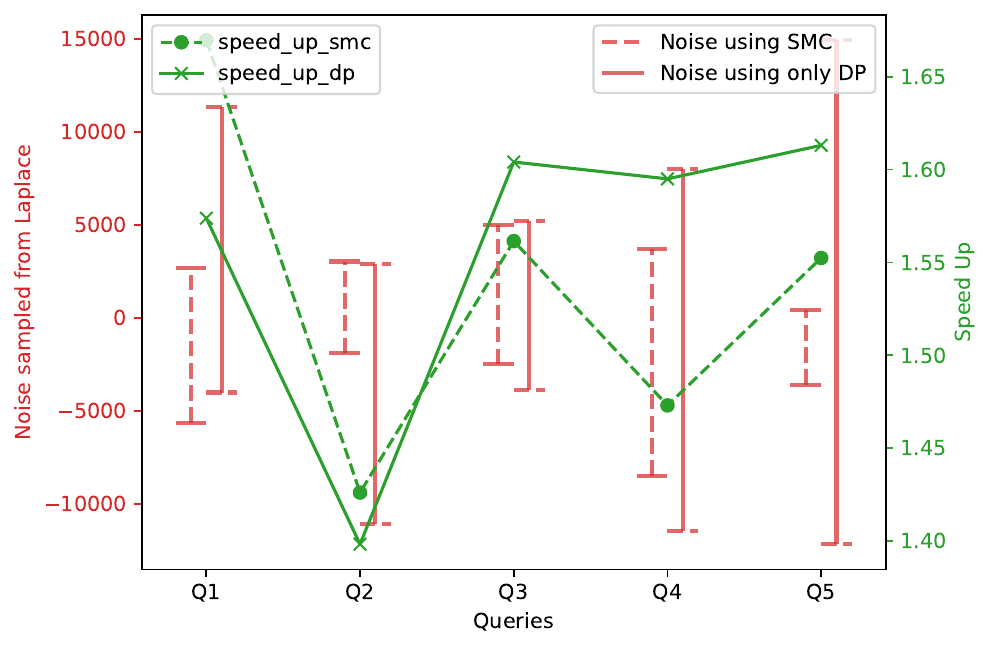}
    \caption{SMC effect on speed-up and accuracy}
    \label{fig:smc-effect}
\end{figure}

The results in Figure \ref{fig:smc-effect} show, for each query, the range of noise sampled using the \textit{Laplace mechanism} for both solutions at each iteration and speed-up. 
We notice in Figure \ref{fig:smc-effect} that using SMC to share only the sensitivity and the local result does not produce significant overhead, which corresponds to the simulation results in Figure \ref{fig:smc_sim}.
Concerning the injected noise, which affects the precision of the query result, the use of SMC allows a more restricted range of perturbation. Meanwhile, if each data provider perturbs its local data without SMC, there could be two cases: (i) the noises from the data providers cancel each other out, or (ii) the noise accumulates.
In the first case, the sum of noises is close to zero because some are positive and others negative, which will help improve accuracy. In the second case, which represents the worst case where most of the noise is positive or negative, the accuracy of the results will be greatly affected. 

Based on the experiment results, a user/data provider can choose the appropriate query execution process (with or without SMC) based on their needs, preferring accuracy over speed-up or vice versa.

\subsection{Resilience to Learning-Based Attacks}
DP prevents membership attacks revealing the presence/absence of an individual in the database. In \cite{cormod}, the author introduced a simple attack that allows the disclosure of an individual's sensitive $SA$ attribute based on anonymized data. 
This attack relies on training a \emph{Naive Bayes Classifier} (NBC) using the results of \verb+COUNT+ queries from a noisy database, and this classifier will be used to predict the value of $SA$ based on a given set of $QI$ (quasi-identifiers) attribute values of an individual. 
In our data model, $SA$ corresponds to one of the dimensions $d_{SA} \in D$, and $QI$ is the  subset  $D_{QI} \subseteq D\setminus 
\{d_{SA}\}$. Given $V_{QI} =\{v_1, ..., v_{|D_{QI}|}\}$ for $D_{QI}$,
a NBC attaches a probability to each possible value $y$ of $d_{SA}$ ($y \in |d_{SA}|$).
The predicted value $\hat{y}$ is the one with the highest probability according to Bayes Theorem \cite{cormod}:
\[
\hat{y} = \underset{y \in |d_{SA}|}{\arg\max} \text{ } P(y) \prod^{|D_{QI}|}_{i=1} P(v_i | y)/P(v_i) 
\]
To make these predictions, the classifier goes through a training phase during which it learns the conditional probabilities using the queries \verb+COUNT(*)+ (or \verb+SUM(Measure)+) issued by the attacker to the database. 
The learned probabilities are saved and later used to make predictions. 
The number of queries $nQueries$ needed is:
\[
nQueries = 1 + ||d_{SA}|| + ||d_{SA}|| \times \sum_{d{QI} \in D_{QI}} ||d{QI}||
\]
which is used to compute the \emph{size} of the database, $P(y)$ and $P(v | y)/P(v)$ for all values and dimensions.
For instance, consider a table \verb+T+ with $10000$ rows and $|d_{SA}|  = [20,.., 60]$ is the dimension for \textit{Age} attribute. To compute $P(Age = 25)$, we use the following \verb+COUNT+ query: 
\verb+SELECT COUNT(*) FROM  T WHERE  25 <= Age <= 25 )/ 10000+.
This huge number of queries can be easily issued to a published database using a DP algorithm with a fixed privacy budget (e.g. PrivBayes\cite{privbayes}), and from which the attacker can infer some knowledge \cite {cormod,usnix}.

However, the database is  \textit{not published} in our system.  As we showed in Section \ref{sec:accounting} the attacker has a limited budget $(\xi > 0,\psi > 0)$, from which each issued query consumes a privacy budget $(\epsilon > 0, \delta > 0)$ based on a sequential composition \ref{th:parallel}. 
Since $nQueries$ can be very large, $\epsilon$ must be very small 
$\epsilon = \xi / nQueries$ and $\delta = \psi / nQueries $, thus losing the utility of query answers. 
An alternative to sequential composition  is \textit{Advanced composition} \cite{book, advenced_compo}, which allows the queries to have a greater budget $\epsilon$ without exceeding $\xi$.  
With the advanced composition,  the budget of each query is: 
$\epsilon = \xi /\left(   2 \times \sqrt{2\times nQueries \times log(\frac{1}{\delta})}\right) \text{ and } \delta = \psi/ nQuesries$. We notice that $\xi /\left(   2 \times \sqrt{2\times nQueries \times log(\frac{1}{\delta})}\right) >  \xi / nQueries$, which means queries have better utility.

To evaluate the resilience of our system against this learning-based attack, we tested both sequential compositions and the two allowed queries \verb+COUNT+ and \verb+SUM+. 
We also considered \emph{parallel composition} which allows multiple attackers to create a coalition, where each of them executes only one query (to maximize utility) and combines it with those of other attackers to train the classifier. The ingredients of our experiments are as follows:
 
\noindent\textbf{Setup:} We used \textit{Adult} dataset with four data providers. We selected $3$ dimensions of our table to be $D_{QI}$ and $1$ dimension to be $d_{SA}$ where $||d_{SA}|| = $100 (i.e. the number of classes for NBC).
We also set $\psi = 10 ^{-6}$ and we varied $\xi$ between $1$ and $100$  since there is no standard value \cite{book, budget_paper}. 

\noindent\textbf{Evaluation:} To assess the quality of the learning attack, we measured the accuracy of the NBC in predicting the value of $SA$ for each row in the original table $accuracy = \frac{ \text{number of correct predictions}}{ \text{total number of predictions}}$.

\begin{table}[h]
    \centering
    \begin{tabular}{|c|c|c|c|c|} \hline 
         &   $\xi = 1$&  $\xi = 20$&  $\xi = 50$& $\xi = 100$\\ \hline 
         Sequential / COUNT&   $<   1\%$&  $<   1\%$&  $<   1\%$& $<   1\%$\\ \hline 
         Sequential / SUM&  $<   1\%$&  $<   1\%$&  $<   1\%$& $<   1\%$\\ \hline 
         Advanced  / COUNT&  $<   1\%$&  $<   1\%$&  $<   1\%$& $<   1\%$\\ \hline 
          Advanced / SUM&  $<   1\%$&  $<   1\%$&  $<   1\%$& $<   1\%$\\ \hline 
         Coalition / COUNT&  $<   1\%$&  $<   1\%$&  $<   1\%$& $<   1\%$\\ \hline 
 Coalition  / SUM& $<   1\%$& $<   1\%$& $<   1\%$&$<   1\%$\\ \hline
    \end{tabular}
    \caption{Inference accuracy based on $\xi$}
    \label{tab:learning_attack}
\end{table}

The results in Table \ref{tab:learning_attack} show that in all scenarios the accuracy is $<1\%$. Since the $SA$ we used had $100$ possible values, this means that the trained  classifier is given similar accuracy as randomly assigning a value for $SA$ in each row. Three reasons can be put forward to explain the failure of the learning-based attack: 
i) our system is interactive (the database is not released) and the budget is limited, thus it is difficult to have good accuracy for large numbers of queries by a single attacker; 
ii) query answers in our system are approximated with random sampling, which will introduce some error; 
iii) the smooth sensitivity has a considerable scale, and in the case of queries that collects small values, the accuracy can be lost even for large values of $\epsilon$.\\

Similar results were obtained when fixing the $\xi =100$ and changing the number of dimensions in $D_{QI}$ from 1, 3, 5 to 8. This shows the resilience of our system in different settings.

\section{Discussion}\label{sec:discussion}
In this section, we discuss the constraints, limits and points of improvement that could be integrated into our solution. In order to approximate the sampling probabilities in Section \ref{sec:approx}, we assumed that the dimensions are independent and that there is no correlation between them.
However, this assumption is not valid in some cases. For example, if an individual's $Age$ is less than $25$, this implies with a high probability that he/she is still studying ($profession = student$). Likewise, if $Age > 65$, the attribute $profession = retired$. When it comes to range queries, capturing and managing these dependencies is non-trivial; so we will leave it for future work.

We also restricted the data providers to using the same value of $S$ in order to approximate the $R$. Otherwise, we cannot compare the $\text{Avg}({\widehat{R}})$ in the allocation phase (Section \ref{sec:allo}). 
To agree on the same $S$, each data provider $\mathbb{S}_i$ can share their true $S_i$ with the others, and they will use then the maximum $S_i$ (which will guarantee that all the $R's$ computed are $\leq 1$). 
The value of $S_i$ itself is not sensitive since it is usually a constant in a database system. But if this is deemed sensitive in a particular case, then data providers can simply share a randomly chosen $S'_i$ value such that: $S_i \leq S'_i \leq S^m_i$ where $S^m_i $ is an upper bound chosen by each data provider (e.g. $S^m_i = 2 \times S_i$).

In our solution, we focused on protecting the intermediate (summary information) and final result from inference attacks with the use of \textit{Differential Privacy}.
However, we have not directly addressed the risks associated with side-channel attacks. It is easy to see that thanks to the collaboration method that we propose, we manage to avoid certain risks mentioned in \cite{doquet}, such as: \textit{memory access models} and \textit {communication volumes } since all data-based computations are performed locally at each data provider and the communication cost is constant and independent of the query. But we have postponed further consideration of this aspect of the problem to dedicated work.

Our solution serves as the first building block towards a more comprehensive solution that handles more complex queries, such as \verb+GROUP-BY+  queries. Integrating such clauses in the SQL query is not so trivial, and adding noise to the final result will not be enough to guarantee privacy \cite{desfontaines2020differentially}. Other aggregations, such as average, standard deviation, and variance, can be derived from SUM and COUNT using the sequential composition of DP. However, to handle other aggregations (such as Min, Max and Mode), different estimators are required. 

Finally, during our evaluation, we built a proof of concept of our solution on \textit{PostgreSQL}. It would be interesting to incorporate it directly into any DBMS, which would further improve our results.

\section{Conclusion}\label{sec:conclusion}
In our study, we introduced a lightweight collaborative approach for online range query approximation in a federated environment. Our experimental results demonstrated the performance improvements our solution is capable of delivering, with processing times improved by up to 8x compared to plain-text execution, while ensuring end-to-end privacy with minimal loss of accuracy.
Our solution uses cluster sampling and query estimation techniques that take into account data distribution to preserve query utility in terms of speed and accuracy. This work lays a solid foundation for future work
to handle more complex queries while maintaining the same level of performance.


\bibliographystyle{ACM-Reference-Format}
\bibliography{Main}

\appendix

\section{Sensitivity summarised information}\label{apn:collab_data}
In order to obtain the allocation (sample size) $s$ based on the inter/intra data provider data distribution, each data provider communicate $Avg(\widehat{R_i})$ and $N^Q$. The sensitivity of the $N^Q$ is straight forward,  given a query $Q$ and  any two neighbouring database $T$ and $T'$ : $\Delta_{N^Q} = \left| N^Q - N'^Q \right| \leq 1 $. Adding/removing and individual at most add/remove a cluster $C$ from $C^Q$.  For the sensitivity of $Avg(\widehat{R_i})$, we need to consider first the sensitivity of a single $R$.

\subsection{Sensitivity $R$}
Given a range query $Q$ and  two neighbouring databases $T$ (with cluster $C$) and $T'$ (with cluster $C'$),  we consider the case where $C'$ has and additional row for an extra individual. that implies:
\begin{equation}
    \begin{array}{ll}
         R = \prod^{d\in D^Q} R^d \text{ and } R' = \prod^{d\in D^Q} (R^d+\frac{1}{S})\\
         \Delta_R = \left| R' - R \right|  = \prod^{d\in D^Q} (R^d+\frac{1}{S}) - \prod^{d\in D^Q} R^d
    \end{array}
\end{equation}
where $D^Q$ is the set of dimensions defining $Q$.
In order to obtain the upper bound of $\Delta_R$, we consider $R' = 1$ which implies $R = (1 - \frac{1}{S})^{\left| D^Q \right|}$ :
\begin{equation}
    \Delta_R = 1 - (1 - \frac{1}{S})^{\left| D^Q \right|}
\end{equation}
Since the values of $\left| D^Q \right|$ and $S$ are publicly known, there no information leak when using $\Delta_R$ based on these values.
The other possible scenarios of neighbouring are : 1) $C'$ has on row less, which will give the same result as the previous one. 2) $C'$ has new/lost individual but only affected the column \textit{"Measure"} of a row by $\pm$ 1, then $\Delta_R =0$. 3) Case where a cluster $C$ wasn't in $C^Q$ in $C'$ has an additional row and his in $C^Q$ (or vice versa), in this case $\Delta_R = \frac{1}{S^{\left| D^Q \right|}}$.
We can prove that $1 - (1 - \frac{1}{S})^{\left| D^Q \right|} \ge \frac{1}{S^{\left| D^Q \right|}}$ :
 
\begin{equation}
    \begin{array}{ll}
         1 - (1 - \frac{1}{S})^{\left| D^Q \right|} = 1 - (1 - \frac{\left| D^Q \right|}{S}+\frac{\left| D^Q \right| \times (\left| D^Q \right| -1)}{2\times S^2} - ...) \\
         1 - (1 - \frac{1}{S})^{\left| D^Q \right|} = \frac{\left| D^Q \right|}{S}-\frac{\left| D^Q \right| \times (\left| D^Q \right| -1)}{2\times S^2} + ...
    \end{array}
\end{equation}
Since $S \gg D$, we can assume $1 - (1 - \frac{1}{S})^{\left| D^Q \right|} \approx \frac{\left| D^Q \right|}{S}$. Then :
\begin{equation}
    1 - (1 - \frac{1}{S})^{\left| D^Q \right|} > \frac{1}{S^{\left| D^Q \right|}} \Leftrightarrow  \frac{\left| D^Q \right|}{S} \ge \frac{1}{S^{\left| D^Q \right|}}
\end{equation}
Which is always true ($S \gg 0 \text{ and } \left| D^Q \right| \ge 1 $)

\subsection{Sensitivity $Avg(\widehat{R})$} \label{apn:avg_R}
The average of $R's$, $\widehat{R}$, of a data provider's set of cluster $C^Q$ is computed as follows :  $Avg(\widehat{R}) = \frac{\sum^{R \in \widehat{R}}R}{N^Q}$. For any two neighbouring databases $T$ and $T'$, there is two cases where the $Avg(\widehat{R})$ is effected: 1)One of the clusters in $C'^Q$ has additional row compared to his counter part in $C^Q$ 2) $C'Q$ has new cluster due to the presence of an individual thus $N'^Q = N^Q + 1$. Which will give:

\begin{equation}
    \Delta _{Avg(\widehat{R})} = \left\{
    \begin{array}{ll}
         &  \left|  \frac{\sum^{R \in \widehat{R}}R}{N^Q} -  \frac{\Delta_R + \sum^{R \in \widehat{R}}R }{N^Q}\right| = \frac{\Delta_R}{N_Q}  \\
         & \left|  \frac{\sum^{R \in \widehat{R}}R}{N^Q} -  \frac{\frac{1}{S^{\left| D^Q \right|}} + \sum^{R \in \widehat{R}}R }{N^Q + 1}\right|
    \end{array}
    \right.
\end{equation}

We can simplify the the second $\Delta_{Avg(\widehat{R})}$ as follows:
\begin{equation}
    \begin{array}{ll}
       \left|  \frac{\sum^{R \in \widehat{R}}R}{N^Q} -  \frac{\frac{1}{S^{\left| D^Q \right|}} + \sum^{R \in \widehat{R}}R }{N^Q + 1}\right| = \left|  \frac{Avg(\widehat{R})}{N^Q + 1} -  \frac{\frac{1}{S^{\left| D^Q \right|}}}{N^Q+1}\right|\\
       \text{ which is :} \leq \frac{1 -\frac{1}{S^{\left| D^Q \right|}} }{N^Q+1} \le \frac{1}{N^Q + 1}
    \end{array}
\end{equation}
$N^Q$ can be replaced by it's smallest possible value $N^{min}$ to maximise $\Delta_{Avg(\widehat(R))}$ : 
We can simplify the the second $\Delta_{Avg(\widehat{R})}$ as follows:
\begin{equation}
    \begin{array}{ll}
       \left|  \frac{\sum^{R \in \widehat{R}}R}{N^Q} -  \frac{\frac{1}{S^{\left| D^Q \right|}} + \sum^{R \in \widehat{R}}R }{N^Q + 1}\right| = \left|  \frac{Avg(\widehat{R})}{N^Q + 1} -  \frac{\frac{1}{S^{\left| D^Q \right|}}}{N^Q+1}\right|\\
       \text{ which is :} \leq \frac{1 -\frac{1}{S^{\left| D^Q \right|}} }{N^Q+1} \le \frac{1}{N^Q + 1}
    \end{array}
\end{equation}
$N^Q$ can be replaced by it's smallest possible value $N^{min}$ to maximise $\Delta_{Avg(\widehat(R))}$ : 

\begin{equation}
\Delta _{Avg(\widehat{R})} \leq \left\{
    \begin{array}{cc}
         & \frac{\Delta_R}{N^{min}}  \\
         & \frac{1}{N^{min} + 1}
    \end{array}
    \right. \implies \Delta _{Avg(\widehat{R})} = max(\frac{\Delta_R}{N^{min}},\frac{1}{N^{min} + 1})
\end{equation}

Since $\Delta_R$ and $N^{min}$ are not sensitives information, they can be used to express the $\Delta_{Avg(\widehat(R)}$.
\section{Sensitivity estimator $\mathbb{E}$}\label{apn_estimator}
For the estimator used to approximate the result of $Q$, we first will give the bound for its global sensitivity, then we show how we bound his local sensitivity.
\subsection{Global sensitivity of the estimator $\mathbb{E}$}
Given a query $Q$, two neighbouring databases $T$ and $T'$ containing cluster $C \in C^Q$ and $C' \in C'^Q $ where $C'$ has an additional row that covers $Q$. Thus both the sampling probability $p$ and $Q(C)$ are effected by this additional row, and we have $p'$ and $Q(C')$, this implies :
\begin{equation}
    \begin{array}{ll}
         \Delta_{\mathbb{E}} = \left| \frac{Q(C)}{p} - \frac{Q(C')}{p'} \right| \text{ where } Q(C') = Q(C)+1  \\
         \Delta_{\mathbb{E}} = \left| \frac{Q(C)}{p} - \frac{Q(C)+1}{p'} \right| = \left| \frac{Q(C)}{p} - \frac{Q(C)+1}{p'} \right| = \frac{1}{p \times p'} \times \left| Q(C)\times (p' - p) - p \right| 
    \end{array}
\end{equation}
Since the $p\times p'$ is in the denominator, we can  minimise it to obtain the $\Delta_{\mathbb{E}}$. Let's consider the case where $C$ contains only one row $R_C = \frac{1}{S^{\left| D^Q \right|}}$ that covers $Q$ and the remain cluster in $C^Q$ fully covers $Q$ so their $R = 1$ which implies :
\begin{equation}\label{eq:unbounded_est_5}
    p = \frac{1/S^{\left| D^Q \right|}}{ N^Q - 1 + 1/S^{\left| D^Q \right|}} = \frac{1}{(N^Q - 1)\times S^{\left| D^Q \right| }+1}
\end{equation}
In $T'$, $C'$ has an additional row so $p'$ is:
\begin{equation}\label{eq:unbounded_est_6}
    p' = \frac{2^{\left| D^Q \right|}/S^{\left| D^Q \right|}}{ N^Q - 1 + 2^{\left| D^Q \right|}/S^{\left| D^Q \right|}} = \frac{2^{\left| D^Q \right|}}{(N^Q - 1)\times S^{\left| D^Q \right| }+ 2^{\left| D^Q \right|}}
\end{equation}
From equations (\ref{eq:unbounded_est_5}) and (\ref{eq:unbounded_est_6}) :
\begin{equation}
\left\{
    \begin{array}{ll}
         & \frac{1}{p \times p'} = \frac{((N^Q - 1)\times S^{\left| D^Q \right| }+1)\times ((N^Q - 1)\times S^{\left| D^Q \right| }+ 2^{\left| D^Q \right|})}{2^{\left| D^Q \right|}}  \\
         & p' - p = \frac{(2^{\left| D^Q \right|} -1)\times (N^Q -1)\times S^{\left| D^Q \right|}}{((N^Q - 1)\times S^{\left| D^Q \right| }+1)\times ((N^Q - 1)\times S^{\left| D^Q \right| }+ 2^{\left| D^Q \right|})}
    \end{array}
    \right.
\end{equation}

Which implies:
\begin{equation}
    \begin{array}{ll}
         \Delta_{\mathbb{E}} = \frac{1}{2^{\left| D^Q \right|}} \times \left| 2^{\left| D^Q \right| -1}\times(N^Q-1)\times S^{\left| D^Q \right|} - 2^{\left| D^Q \right|} \right| \\
        \Delta_{\mathbb{E}} = \frac{(N^Q-1) \times S^{\left| D^Q \right|} }{2} -1
    \end{array}
\end{equation}

This results shows that the sensitivity of this statistical estimator is very large and unbounded, if its result should be protected another alternative is mandatory.
\subsection{Smooth sensitivity of the estimator $\mathbb{E}$}\label{apn:decide}
To compute the smooth upper bound of the $LS_{\mathbb{E}}$, we considered four possible scenarios of neighbouring $T$ and $T'$:
\begin{enumerate}
    \item $LS_{\mathbb{E}}^1 = \left| \frac{Q(C)}{p} - \frac{Q(C)}{p'}  \right|$ and $p' = \frac{R}{\Delta_R + \sum^{R \in \widehat{R}}R}$ : another cluster gained a row
    \item $LS_{\mathbb{E}}^2 = \left| \frac{Q(C)}{p} - \frac{Q(C)}{p'}  \right|$ and $p' = \frac{R}{1/S^{\left| D^Q \right|} + \sum^{R \in \widehat{R}}R}$ : new cluster added to $C^Q$
    \item $LS_{\mathbb{E}}^3 = \left| \frac{Q(C)}{p} - \frac{Q(C')}{p'}  \right|$ and $p' = \frac{R+\Delta_R }{1/S^{\left| D^Q \right|} + \sum^{R \in \widehat{R}}R}, Q(C') = Q(C) + 1$ : the cluster gained a row.
    \item $LS_{\mathbb{E}}^4 = \left| \frac{Q(C)}{p} - \frac{Q(C')}{p}  \right|$ and $Q(C') = Q(C) + 1$ : the cluster gained an individual $\pm$ 1 in a measure and not a new row.
\end{enumerate}

Our goal is to find the biggest one of these distances. we can quickly notice that $LS_{\mathbb{E}}^1 > LS_{\mathbb{E}}^2$ since $\Delta_R > \frac{1}{S^{\left| D^Q \right|}}$. And we have $LS_{\mathbb{E}}^4 > LS_{\mathbb{E}}^3$ because this constraint is always true:
\begin{equation}
    \begin{array}{ll}
         \frac{Q(C')/p}{Q(C')/p'} = \frac{p'}{p} = \frac{R + \Delta_R}{R} >1 \\
         \implies  LS_{\mathbb{E}}^4 > LS_{\mathbb{E}}^3
    \end{array}
\end{equation}

Between  $LS_{\mathbb{E}}^1$ and $LS_{\mathbb{E}}^4$, we need to find which is the bigger distance and under what conditions:
\begin{equation}
    \begin{array}{ll}
         \frac{Q(C)/p'}{Q(C')/p} = \frac{Q(C)\times(\Delta_R + \sum^{R \in \widehat{R}}R )}{R} \times \frac{R}{(Q(C)+1)\times \sum^{R \in \widehat{R}}R}  \\
         \frac{Q(C)/p'}{Q(C')/p} = \frac{Q(C)}{Q(C)+ 1} \times \frac{\Delta_R + \sum^{R \in \widehat{R}}R )}{ \sum^{R \in \widehat{R}}R )} \\
         \frac{Q(C)/p'}{Q(C')/p} > 1 \implies Q(C) > \frac{\sum^{R \in \widehat{R}}R}{\Delta_R}
    \end{array}
\end{equation}

In conclusion, only $LS_{\mathbb{E}}^1$ and $LS_{\mathbb{E}}^4$ need to be used in order to compute the smooth sensitivity, and for each cluster there is only one that dominated the other based if $Q(C) > \frac{\sum^{R \in \widehat{R}}R}{\Delta_R}$. Both distances can be simplified as follows :

\begin{equation}
    LS_{\mathbb{E}}^ = \left\{
    \begin{array}{ll}
         & \frac{Q(C)\times \Delta_R}{R} \text{  from } LS_{\mathbb{E}}^1  \\
         & \frac{1}{p} \text{  from } LS_{\mathbb{E}}^4
    \end{array}
    \right. 
\end{equation}

\subsection{Bound the k for the smooth sensitivity}\label{apn:bound_k}
Since we have our distances $LS_{\mathbb{E}}^k$ are ascending function and $e^{-k \beta}$ is an exponential decay function, and since we are looking for the $max_{k=0,1,...} e^{-k \beta} \times LS_{\mathbb{E}}^k $. To find the stopping point of $k$, we need to find  where the this product starts decaying. In other words, we find the $k$ such that : 
\begin{equation}
    e^{-k \beta} \times LS_{\mathbb{E}}^k  < e^{-(k-1)\beta} \times LS_{\mathbb{E}}^{k-1} \implies \frac{LS_{\mathbb{E}}^{k-1}}{LS_{\mathbb{E}}^k} > e^{- \beta}
\end{equation}
For $LS_{\mathbb{E}}^k$ based on scenario 1:
\begin{equation}
    \frac{LS_{\mathbb{E}}^{k-1}}{LS_{\mathbb{E}}^k} = \frac{(k-1)\times Q(C) \times \Delta_R}{R} \times \frac{R}{k\times Q(C) \times \Delta_R} = \frac{k-1}{k}
\end{equation}
For $LS_{\mathbb{E}}^k$ based on scenario 4:
\begin{equation}
    \frac{LS_{\mathbb{E}}^{k-1}}{LS_{\mathbb{E}}^k} = \frac{k-1}{p} \times \frac{p}{k} = \frac{k-1}{k}
\end{equation}
So for both our distances, the smooth upper bound is reached when:
\begin{equation}
    \frac{k-1}{k} > e^{-\beta} \implies k > \frac{1}{1 - e^{-\beta}}
\end{equation}
Where $\beta = \frac{\epsilon}{2 \times\ln(2/\delta)} $.
Based on the $(\epsilon,\delta)$ budget  we set for the estimator, we will obtain  our exact upper bound for the $k$. this shows that our process terminates and don't run indefinitely.

\end{document}